\documentstyle[pre,aps,multicol,eqsecnum,epsf,rotate]{revtex}
\newenvironment{mcols}{\begin{multicols}{2}}{\end{multicols}}
\begin{document}
\draft
\tighten
\title
{\LARGE Critical Behavior of the Supersolid transition
in Bose-Hubbard Models}
\author{Erwin Frey}
\address{
Institut f\"ur Theoretische Physik,
Physik-Department der Technischen Universit\"at M\"unchen, \\
James-Franck-Stra\ss e, D-85747 Garching, Germany}
\author{Leon Balents}
\address{
Institute for Theoretical Physics, University of California, \\
Santa Barbara, CA 93106-4030}
\date{\today}
\maketitle


\begin{abstract}
  
  We study the phase transitions of interacting bosons at zero temperature
  between superfluid (SF) and supersolid (SS) states.  The latter are
  characterized by simultaneous off-diagonal long-range order and broken
  translational symmetry.  The critical phenomena is described by a
  long-wavelength effective action, derived on symmetry grounds and verified by
  explicit calculation.  We consider two types of supersolid ordering:
  checkerboard (X) and collinear (C), which are the simplest cases arising in
  two dimensions on a square lattice.  We find that the SF--CSS transition is
  in the three-dimensional XY universality class.  The SF--XSS transition
  exhibits non-trivial new critical behavior, and appears, within a
  $d=3-\epsilon$ expansion to be driven generically first order by
  fluctuations. However, within a one--loop calculation directly in $d=2$ a
  strong coupling fixed point with striking ``non-Bose liquid'' behavior is
  found. At special isolated multi-critical points of particle-hole symmetry,
  the system falls into the 3d Ising universality class.

\end{abstract}



\begin{mcols}
\section{Introduction}

Because of the dominant role played by transport measurements,
conducting to insulating transitions merit special attention in
condensed matter systems.  In electronic systems, examples include the
Anderson\cite{Anderson}\ and Mott\cite{Mott}\
transitions\cite{AMreview}.  In bosonic systems, interest has focused
on the disorder-dominated regime: the so-called superfluid--insulator
transition\cite{SFI}.  Surprisingly, the somewhat simpler problem of
superfluid-insulator transitions in {\sl clean} bosonic systems is
less well studied.  We will see that in such systems, the path from
the superfluid to the insulating state generically occurs via an
intermediate ``supersolid'' phase.

Recent work has endeavored to remedy this deficiency through numerical
and mean-field studies of disorder--free interacting boson
models\cite{BFS93,RS93,Wag94,vOtt94,vOW94,BSZ95,KZB95,vOW95}.  Such
models are experimentally relevant both for helium on a periodic
substrate, and for two dimensional Josephson junction arrays (see
below).

\subsection{Model Hamiltonians}

In the case of Helium with a strong substrate potential, a
tight-binding description is appropriate.  This leads to the
Bose-Hubbard Hamiltonian
\begin{eqnarray}
  H = 
  &&-t \sum_{\langle ij \rangle} 
    \left( a_i^\dagger a_j + a_j^\dagger a_i  \right) - \mu  \sum_i n_i
  \\ \nonumber 
  &&+ U_0 \sum_i n_i^2 + U_1 \sum_{\langle ij \rangle} n_i n_j
    + U_2 \sum_{\langle \langle ik \rangle \rangle} n_i n_k \, ,
\label{bose_hubbard_hamiltonian}
\end{eqnarray}
where $a_i^\dagger$ and $a_i$ are bosonic creation and annihilation
operators obeying $[a_i,a_j^\dagger] = \delta_{ij}$, and $n_i =
a_i^\dagger a_i$.  The parameter $t$ is a hopping matrix element,
$\mu$ is the chemical potential, and $U_0$, $U_1$, and $U_2$ are
on--site, nearest--neighbor, and next--nearest--neighbor interactions,
respectively.

For a Josephson junction array, the relevant parameters are the charge $n_i$
residing on each superconducting island and its phase $\phi_i$.  Including
capacitive and Josephson couplings leads to the quantum phase Hamiltonian
\begin{eqnarray}
  H & = & -J \sum_{ \langle ij \rangle } \cos (\phi_i - \phi_j)
          - \mu \sum_i n_i \nonumber \\
  &&+ U_0 \sum_i n_i^2 + U_1 \sum_{\langle ij \rangle} n_i n_j
    + U_2 \sum_{\langle \langle ik \rangle \rangle} n_i n_k \, ,
\label{phase_hamiltonian}
\end{eqnarray}
where the chemical potential is tuned by the applied gate
voltage\cite{BFK92}.  The conjugate variables satisfy the commutation
relation $[\phi_i, n_j] = i\delta_{ij}$, and the strength of the
Josephson phase coupling is characterized by the hopping matrix
element $J$.

Finite--range interactions allow for non-trivial charge ordering with
fillings of less than one particle per site.  If the interactions are
very strong, these are simply commensurate crystals.  For weak
interactions, the ground state is, by contrast, a featureless
superfluid (phase ordered) state.  In the intermediate regime, as
discussed below, one may obtain coexisting crystallinity (charge
order) and superfluidity (phase or off--diagonal--long--range--order
(ODLRO)).  Such a state is termed a supersolid.


\subsection{Mean field phase diagram and Monte Carlo simulations} 

The phase diagrams resulting from the above models for interacting
bosons have been determined within mean-field theory
\cite{BFS93,RS93,Wag94,vOW95,MT70,LF73}\ and quantum Monte Carlo
simulations\cite{Wag94,vOtt94,vOW94,BSZ95,KZB95,vOW95,BSZ90,NFS94,KT91,SWG92}.

A representative (mean--field) phase diagram is shown in
Fig.~\ref{phase_diagram}, for the case of the quantum phase model with
$U_2=0$ and $U_1/U_0 = 1/5$.  The mean-field theory gives four
different phases: (1) superfluid (SF), (2) Mott insulator (MI), with
homogeneous density, (3) checkerboard solid (X), or Mott-insulating
checkerboard charge-density wave, and (4) compressible checkerboard
supersolid (XSS or SS).  For $U_2 \geq U_1$, ``collinear'' solid (C)
and supersolid (CSS) phases replace the X and XSS states near
half--filling\cite{vOW95}.  The supersolid (SS) to crystal (X)
transition is discussed in Refs.~\cite{FNF94,BN95}.  It corresponds to
the binding of vacancy or interstitial loops, and is expected to be
mean-field like (if not first order) in two dimensions.  Here, we are
interested in the SS-superfluid (SF) transition, which is more
analogous to a melting transition.  To study it, we will employ the
techniques of Landau theory and the renormalization group (RG).

\subsection{Results}

Our results are summarized below.  The CSS--SF transition is in the
$d+1$--dimensional XY universality class, with dynamical scaling
exponent $z=1$.  Moreover, the critical fluctuations lead to a
reduction of the second sound velocity near the transition point.
This reduction is, however, finite, so a conventional description of
the superfluid properties applies within the critical regime.

The XSS--SF transition is more interesting.  At a particular value of
the chemical potential in the middle of the XSS lobe, the system has
asymptotic particle-hole symmetry.  At this multi-critical point, the
transition is $d+1$--dimensional Ising--like ($z=1$), and again the
superfluid properties are conventional.  Away from this chemical
potential, however, the superfluid and critical properties become {\sl
strongly coupled}.  Within a RG analysis in $d=3-\epsilon$ one finds a
complicated crossover regime leading at low energies to ``runaway
flows''. The absence of a stable fixed point to first order in
$\varepsilon$ is often interpreted as signaling a fluctuation--driven
first order transition. This conclusion must, however, be reached with
some caution.  Cases exist in which one finds runaway RG flows in an
$\epsilon$ expansion, but the physical system (for $\epsilon$ of order
one) has a continuous critical point.  The most well known of these,
the normal--superconducting (NS) transition in zero field\cite{NSzf},
is actually quite similar to this problem in that the order parameter
is coupled to ``spectating'' gapless (in the NS case, gauge) degrees
of freedom.

To address this question, we apply a fixed dimension RG approach
directly in $d=2$.  The issue is inherently a strong-coupling one,
outside of the range of applicability of a strict
$\epsilon$-expansion.  The fixed-dimension RG, while not controlled in
the sense of an $\epsilon$-expansion, does allow at least an
approximate treatment of the features associated with a putative
``strong coupling'' fixed point.  In this situation, the superfluid
properties are unconventional.  This scenario is the bosonic analog of
quantum critical modifications of electronic properties recently
discussed by Millis {et\ al.}\cite{Millis}.  It is thereby natural to
denote such anomalous superfluid behavior as ``non--Bose liquid.''
The non--Bose liquid phase is characterized, for example, by anomalous
dispersion $\omega \sim k^z$ for the strongly-coupled phase and CDW
fluctuation modes, with $z=4/3$ within the fixed-dimension RG
approximation.

\subsection{Organization}

The remainder of the paper is organized as follows. In section II we give a
derivation of a field theory for the order parameter fields on the basis of
general symmetry arguments. It is shown that the field theory can be
transformed into a ``phonon'' representation, whose form is closely related to
field theories describing the influence of elastic degrees of freedom on the
critical behavior of magnetic systems. Section III serves to connect the
correlation functions of the field theoretic model with physical observables. A
short discussion of the mean--field phase diagram is given in section IV. The
superfluid--collinear supersolid and superfluid--checkerboard supersolid
transition are analyzed in sections V and VI using standard methods of
Wilson's renormalization group theory. More sophisticated field--theoretic
techniques are used in section VII to analyze the complete crossover 
and to calculate effective critical exponents. In the final section we give a
short summary of our results and discuss some open problems.

The appendices contain additional details and explicit calculations to
supplement those in the main text.  A Hubbard--Stratonovich derivation of the
long-wavelength effective critical action is given in Appendix
\ref{decoupling_app}.  Appendix \ref{loop_appendix}\ calculates the
momentum-dependent $\Phi^2$ terms via an exponential cut-off scheme, needed to
justify the calculation in section \ref{SF_X_transition}.  Appendix
\ref{app_pert}\ comprises the Feynman diagrams and the corresponding analytical
expressions used in the field--theoretic RG.


\section{Landau theory}

To study the critical phenomena associated with the transition from a
superfluid into a supersolid it is useful to rewrite the original Bose-Hubbard
Hamiltonian for lattice bosons in terms of a field theory for the order
parameter fields.  A formal procedure for deriving such a field theory is given
in appendix~\ref{decoupling_app}. Here, we instead use general symmetry
arguments to determine the most general possible action near the critical
point, and hence describe the universal properties of the phase transition.

This procedure is the quantum analog of Landau's theory of thermal melting.
Within this theory, Landau originally argued that such transitions must be
first order.  As pointed out in Ref.\cite{BN95}, the presence of the periodic
substrate invalidates the original argument and continuous melting is in
principle allowed.  Here we study in detail the low--order commensurate SS
states occurring near 1/2-filling, where such substrate effects are strongest.

There are two sorts of states which arise naturally in a model with up to
next--nearest--neighbor interactions.  If nearest--neighbor interactions
dominate, the ground state is a checkerboard SS, while next--nearest--neighbor
interactions favor striped (or collinear) SSs \cite{BSZ95}.  The order
parameters distinguishing these states are simply Fourier components of the
density $n({\bf x})$:
\begin{equation}
  n({\bf x}) = n_0 + (-1)^{\sum_i x_i}\Phi + \sum_i(-1)^{x_i} \psi_i.
\label{opdefs}
\end{equation}
Here $\Phi$ and the $\bbox{\psi} \equiv \{\psi_1,\cdots,\psi_d\}$ are
checkerboard and collinear order parameters, respectively.  The form of the
action is determined by the transformation properties of these fields.  The
important symmetries are translations by one lattice constant, reflections in a
codimension one subspace, and $\pi/2$ rotations within an arbitrary plane.  The
transformation properties of $\Phi$ and $\bbox{\psi}$ under these operations
are summarized in table \ref{symmetrytable}.  The most general local action
consistent with these symmetries is
\begin{eqnarray}
  S_0 & = &
     \int d{\bf x} d\tau
     \Biggl\{
             \frac{1}{2}\left( \frac{1}{c}
             \partial_\tau \Phi \right)^2 +
             \frac{1}{2}\left( \bbox{\nabla} \Phi \right)^2
             + \frac{t}{2} \Phi^2 + \frac{u}{4!} \Phi^4 \nonumber \\
             && + \frac{1}{2}\left| \frac{1}{{c'}}
             \partial_\tau \bbox{\psi} \right|^2 +
             \frac{1}{2}\left| \bbox{\nabla} \bbox{\psi} \right|^2
             + \frac{t'}{2}|\bbox{\psi}|^2 \nonumber \\
             &&+ \frac{u'}{4!}|\bbox{\psi}|^4 - \frac{v'}{4!} 
                \sum_i \psi_i^4 - w \Phi \prod_i \psi_i
     \Biggr\} \, ,
\label{critical_action}
\end{eqnarray}
where the coordinates and fields are rescaled such that scale of the spatial
gradients is fixed to be $1$. The control parameters $t$ and $t'$ measures the
distances from the critical points, and $c$ and $c'$ are ``velocities''
characterizing the crystalline order (in an incommensurate crystal phase, $c$
would be proportional to the bare phonon velocity).

{\it Phase fluctuations.} Eq.(\ref{critical_action})\ assumes a {\sl local}
form of the action.  Additional non--local interactions arise due to
interactions with long wavelength fluctuations of the superfluid phase,
$\theta$.  As usual, $\theta$ is governed by the action
\begin{equation}
 S_1 = \frac{\tilde{\rho}_s}{{2m^2}}\int \! d{\bf x}d\tau \bigg\{
            \left(\frac{1}{v} \partial_\tau\theta\right)^2 +
            (\bbox{\nabla}\theta)^2\bigg\},
\label{phaseaction}
\end{equation}
where $m$ is the atomic mass, and $\tilde{\rho}_s$ and $v$ are the ``bare''
superfluid density and velocity, respectively.  We have taken ${\tilde \rho}_s$
to be a {\sl mass} density, and work in units such that $\hbar = 1$.  The most
relevant (near the critical point) coupling to the spatial order parameters
allowed by the time-reversal and $U(1)$ symmetries is
\begin{equation}
  S_2 =  \int \! d{\bf x}d\tau \bigg\{ i \sigma \partial_\tau
                     \theta \Phi^2 + i \sigma' \partial_\tau\theta
                      |\bbox{\psi}|^2 \bigg\}.
\label{ptcoupling}
\end{equation}
Note that the factors of $i$ in Eq.(\ref{ptcoupling})\ are dictated by the
requirement (charge conjugation) that $S \rightarrow S^\dagger$ when $a
\rightarrow a^\dagger$.

{\it ``Phonon'' representation}.  The partition function has an alternative,
purely real, representation, which is related to a {\sl classical} statistical
mechanical system in $d+1$ dimensions.  To take advantage of insights gained in
this area, we now develop an alternative representation of the phase
fluctuations.  For this section, we focus for simplicity on the checkerboard
order, and choose $\bbox{\psi}=\bbox{0}$.  Performing a Hubbard--Stratonovich
transformation then results in the replacements
\begin{eqnarray}
  \frac{\tilde{\rho}_s}{{2m^2}}(\bbox{\nabla}\theta)^2  & \rightarrow &
               \frac{1}{{2\tilde{\rho}_s}}J^2 + \frac{i}{m}
               \bbox{J}\cdot\bbox{\nabla}\theta, \\
  \frac{\tilde{\rho}_s}{{2m^2v^2}}(\partial_\tau\theta)^2  + i
               \sigma\partial_\tau \theta \Phi^2 & \rightarrow &
               \frac{v^2}{{2\tilde{\rho}_s}}\left(J_0 -
               m\sigma\Phi^2\right)^2 \nonumber\\
               & & + \frac{i}{m} J_0\partial_\tau\theta.
\label{HStransform}
\end{eqnarray}
The auxiliary $4$-vector field $J^\mu$ has the physical interpretation of the
Euclidean super-current.  Indeed, the original phase field may now be
integrated out in favor of the continuity constraint
\begin{equation}
   \partial_\tau J_0 + \bbox{\nabla}\cdot\bbox{J} = 0.
\label{continuity}
\end{equation}
Eq.(\ref{continuity})\ is solved by introducing the ``phonon''--like field
$\bbox{w}$, according to
\begin{eqnarray}
  J_0 & = & - \frac{\sqrt{\tilde{\rho}_s}}{c}\bbox{\nabla}\cdot\bbox{w},
                \nonumber \\
  \bbox{J} & = & \frac{\sqrt{\tilde{\rho}_s}}{c} \partial_\tau
  \bbox{w}.
\label{wdefinition}
\end{eqnarray}
The final form of the action, expressed in terms of $\Phi$ and $\bbox{w}$, is
then
\begin{eqnarray}
  S = 
  && \int d{\bf x} d\tau 
     \left\{
     \frac{1}{2} \left( 
                  t \Phi^2 + \left( \frac{1}{c} \partial_\tau \Phi \right)^2+
                  \left( \bbox{\nabla} \Phi \right)^2
                 \right) +
     \frac{\tilde{u}}{4!} \Phi^4
     \right\} \nonumber \\ 
  && + \int d{\bf x} d\tau 
     \left\{
     \frac{1}{2} \kappa \left( \bbox{\nabla} \cdot {\bf w}  \right)^2 + 
     \frac{1}{2}  \left( \frac{1}{c} \partial_\tau {\bf w} \right)^2
     \right\} \nonumber  \\
  && + \int d{\bf x} d\tau 
     \left\{
     \frac{g}{2}  \left( \bbox{\nabla} \cdot {\bf w}  \right) \Phi^2
     \right\} \, ,
\label{effective_action}
\end{eqnarray}
where the coefficients $\tilde{u} = u + 12 (v/c)^2
m^2c^2\sigma^2/\tilde{\rho}_s$, $\kappa = (v/c)^2$, and $g=2(v/c)^2
mc\sigma/\sqrt{\tilde{\rho}_s}$.

Closely related models have been studied in the context of the influence of
elastic degrees of freedom on the behavior of magnetic systems near a critical
point \cite{CIsing,dML76,Nat77}. Depending on the number $n$ of spin
components, anisotropy of the elastic constants and the symmetry of the
coupling between the elastic deformations and the order parameter fields one
finds different stability scenarios for the various fixed points of the
renormalization group recursion relations. In the case of an isotropic coupling
between the order parameter field and the phonon field the value of the
specific heat exponent $\alpha$ is crucial for the character of the critical
behavior. For $\alpha > 0$ all fixed points are found to be unstable against
anisotropic perturbations of the elastic degrees of freedom. For $\alpha < 0$
the elastic degrees of freedom, isotropic as well as anisotropic, are
irrelevant and the critical behavior is that of the rigid model. The effect of
anisotropic couplings between the order parameter and the phonon fields has
been considered in Ref.~\cite{Nat77}. It is concluded that a first--order
transition is to be expected for the Ising, anisotropic XY and Heisenberg model
independent of the external conditions. In view of this earlier work it seems
likely that also the above model will exhibit a topology of the flow diagram
which is similar to the magnetic models and which is commonly interpreted as a
signature of a fluctuation driven first--order phase transition. As we will see
from the calculations in sections \ref{SF_Coll_Transition}--\ref{dim_reg} this
is actually the case. However, as will also be discussed in the following
sections, this conclusion has to be reached with some care.


\section {Physical Quantities and Exponents}
\label{physical_quantities}

Near the transition point, the development of the ordered phase is reflected by
the scaling behavior of the structure function.  We define an Euclidean
density--density correlation function
\begin{equation}
S_{\rm E}({\bf x},\tau) \equiv \langle
n(\bbox{x},\tau)n(\bbox{0},0) \rangle_C,
\label{SEdef}
\end{equation}
where the subscript $C$ denotes the cumulant expectation value (connected
Greens function).  Substitution of Eq.(\ref{opdefs})\ exposes the connection to
the order parameters in position space:
\begin{eqnarray}
   S_{\rm E}(\bbox{x},\tau) & \sim & \langle
     \Phi(\bbox{x},\tau)\Phi(\bbox{0},0)\rangle_C (-1)^{\sum_i x_i}
     \nonumber \\
   & & + \sum_{ij} \langle \psi_i(\bbox{x},\tau) 
       \psi_j(\bbox{0},0)\rangle_C (-1)^{x_i},
\label{SErealspace}
\end{eqnarray}
where non--singular smooth contributions have been dropped.  Scaling theory
gives the form of these connected correlations in the critical regime,
\begin{eqnarray}
  \langle \Phi(\bbox{x},\tau)\Phi(\bbox{0},0)\rangle_C & \sim &
  \xi^{2-d-z-\eta} f_\pm(x/\xi,\tau/\xi^z) , \\
  \langle \psi_i(\bbox{x},\tau) \psi_j(\bbox{0},0)\rangle_C & \sim &
  \tilde{\xi}^{2-d-\tilde{z}-\tilde{\eta}}\tilde{f}_{\pm,ij}(x/\tilde{\xi},
  \tau/\tilde{\xi}^{\tilde{z}}),
\label{scalingeqns}
\end{eqnarray}
where $\xi$ and $\tilde{\xi}$ are the correlation lengths for the ordering
transitions of the $\Phi$ and $\bbox{\psi}$ fields, respectively, $z$
($\tilde{z}$) and $\eta$ ($\tilde{\eta}$) are the dynamical scaling and
anomalous dimension exponents, and $f_\pm$ ($\tilde{f}_\pm$) are universal
scaling functions (the $\pm$ signs indicate whether the system is above or
below the critical point).  Additionally, there are correlation length
exponents $\nu$ and $\tilde{\nu}$ which relate the correlation length to the
deviation from the critical point, defined by
\begin{eqnarray}
\xi & \sim & t^{-\nu} , \nonumber \\
\tilde{\xi} & \sim & {t'}^{-\tilde{\nu}}. 
\label{correlation_lengths}
\end{eqnarray}
The structure function is obtained by Fourier transformation of $S_{\rm
  E}(\bbox{x},\tau)$.  The static (equal time) correlator is thus
\begin{eqnarray}
  S_0(\bbox{q}) & \equiv & \;\; \sum_{\bbox{x}} S_{\rm
  E}(\bbox{x},\tau=0)e^{i\bbox{q}\cdot\bbox{x}} \nonumber \\
  & \sim & \;\; \xi^{2-z-\eta}g_\pm[(\bbox{q}-\bbox{\pi})\xi] \nonumber \\
  & & \;\; + \sum_{ij}
  \tilde{\xi}^{2-\tilde{z}-\tilde{\eta}}\tilde{g}_{\pm,ij}[(\bbox{q} -
  \pi\hat{\bbox{e}}_i)\tilde{\xi}],
\label{static_sf}
\end{eqnarray}
where $\bbox{\pi} = (\pi,\ldots,\pi)$, $\hat{\bbox{e}}_i$ is a unit vector in
the $x_i$ direction, and $g_\pm$ and $\tilde{g}_{\pm,ij}$ are scaling
functions.  The full dynamical structure factor also has a scaling form,
\begin{eqnarray}
   S_{\rm E}(\bbox{q},\omega) & \equiv & \;\; \sum_{\bbox{x}} \int \!
   d\tau \, S_{\rm
   E}(\bbox{x},\tau)e^{i\bbox{q}\cdot\bbox{x}+i\omega\tau} \nonumber \\
   & \sim & \;\; \xi^{2-\eta}h_\pm[(\bbox{q}-\bbox{\pi})\xi,\omega\xi^z]
   \nonumber \\ 
   & & \;\; + \sum_{ij}
   \tilde{\xi}^{2-\tilde{\eta}}\tilde{h}_{\pm,ij}[(\bbox{q} -
   \pi\hat{\bbox{e}}_i)\tilde{\xi},\omega\tilde{\xi}^{\tilde{z}}],
\label{Euclidean_sf}
\end{eqnarray}
where again $h$ and $\tilde{h}$ are scaling functions (simple relations exist
between $f$,$g$ and $h$, and may be easily derived by Fourier transformation of
Eqs.(\ref{scalingeqns})--(\ref{Euclidean_sf})).

Somewhat simpler results are obtained in the isotropic, ballistic limit, when
$z=1$ and $\tilde{f}_{\pm,ij} \propto \delta_{ij}$.  Focusing on the behavior
near $\bbox{q} = \bbox{\pi}$, the dynamic structure function then takes the
form
\begin{equation}
   S_{\rm E} \sim \xi^{2-\eta}{\cal S}
   [(|\bbox{q}-\bbox{\pi}|^2c^2+\omega^2)\xi^2]. 
\end{equation}
A similar form holds near $\bbox{q} = \hat{\bbox{e}}_i$.  At small arguments,
we expect $g$ to become approximately Lorentzian.  After analytic continuation
($i\omega \rightarrow \omega + i \epsilon$), this implies the existence of a
massive mode with $\omega \sim \pm \sqrt{\xi^{-2} +
  |\bbox{q}-\bbox{\pi}|^2c^2}$ for small $q$.  At the critical point,
regularity of $S_{\rm E}$ requires ${\cal S}(\chi) \sim 1/\chi^{(2-\eta)/2}$,
and thus $S_{\rm E} \sim 1/(|\bbox{q}-\bbox{\pi}|^2c^2+\omega^2)^{(2-\eta)/2}$.
For $\eta>0$, this implies the behavior
\begin{equation}
{\rm Im} S_{\rm ret.}(\bbox{q},\omega) 
\sim \frac{{\sin(\pi\eta/2)}}
          {(\omega^2 - |\bbox{q}-\bbox{\pi}|^2c^2)^{1-\eta/2}}
\theta[|\omega|-|\bbox{q}-\bbox{\pi}|c ],
\label{retarded_function}
\end{equation}
and thus a continuum of modes with linear dispersion.

In addition to these quantities, the critical fluctuations of the checkerboard
order parameter modify the behavior of the superfluid.  To study these
modifications, it is useful to define the current--current correlation function
\begin{equation}
   D_{ij}(\bbox{x},\tau) \equiv \langle J^i(\bbox{x},\tau)
J^j(\bbox{0},0) \rangle,
\label{current_current}
\end{equation}
where $\bbox{J} = (\tilde{\rho}_s/m)\bbox{\nabla}\theta$ is the super-current.
By introducing an infinitesimal generating field into
Eqs.(\ref{critical_action})--(\ref{ptcoupling})\ and differentiating the final
effective action twice, one easily obtains the expression
\begin{equation}
  D_{ij}(\bbox{x},\tau) = \tilde{\rho}_s \bigg[ \frac{1}{c^2} \langle
                \partial_\tau w^i(\bbox{x},\tau)\partial_\tau
                w^j(\bbox{0},0)\rangle - \delta^{ij}\bigg]
\label{D_formula}
\end{equation}
for $D$ in the phonon representation.

The superfluid density $\rho_s$ and second-sound velocity $v_s$ may be
extracted from the long wavelength form
\begin{equation}
D_{ij}(\bbox{k},\omega) \sim \rho_s \frac{{k_i k_j}}{{k^2 +
\omega^2/v_s^2}},
\label{Dform}
\end{equation}
for $k,\omega \rightarrow 0$.

\section{Mean Field Theory}
\label{mean_field_theory}

Eq.(\ref{critical_action})\ exhibits four phases in mean-field theory.  In
the superfluid (SF) phase there is no spatial order, so $\Phi=0$ and
$\bbox{\psi} = \bbox{0}$.  The checkerboard (X) state has $\phi \neq 0$ but
$\bbox{\psi} = \bbox{0}$.  A collinear (C) or striped phase occurs with
$\Phi=0$ but $\bbox{\psi} \propto \hat{\bbox{e}}_i$.  Lastly, there is a mixed
(M) phase in which no symmetries persist, so $\Phi \neq 0$ and $\bbox{\psi}
\neq \bbox{0}$ and does not point along a symmetry axis of the lattice.

We have been unable to determine the detailed geometry of the mean--field phase
diagram.  However, we expect the general topology shown in
Fig.\ref{MF_phase_diagram_fig}.  For large $t$ and $t'=0$, there is a critical
line separating the SF and C phases, while for large $t'$ and $t=0$, a
different phase boundary separates the SF and X states.


\section{Superfluid--Collinear SS Transition}
\label{SF_Coll_Transition}

For large $t$, the checkerboard order parameter $\Phi$ is strongly suppressed
and plays no role in the SF-C transition.  We therefore consider here
Eq.(\ref{critical_action})\ with $\Phi=0$ and thereby neglect $w$ (indeed,
integrating out the massive $\Phi$ field leads only to a small renormalization
of $u'$ and $v'$).  For concreteness, we now also focus directly on the case
$d=2$, in which $\bbox{\psi}$ is a two-component field.

In the absence of coupling to the superfluid phase mode (i.e. $\sigma'=0$), the
action is identical to the classical free energy of an XY ferromagnet with
cubic (square) anisotropy $v'$ in $D=2+1=3$ dimensions.  For positive $v'$ (as
assumed here), the collinear state is favored for $t' < 0$ in mean-field
theory.  To study the critical properties for $D=3$, however, we must include
the effects of fluctuations.

Such fluctuations have been studied (with $\sigma'=0$) by many authors
\cite{Wal73,Aha73,KW73,BGZ74,NT75,LP75,Rud78,IA81,Aha86}.  There is a general
agreement between epsilon expansion and direct RG calculations as well as
experimental observations~\cite{Aha86} that cubic anisotropy is in fact an
{\sl irrelevant} perturbation on the XY fixed point in three dimensions, with
RG eigenvalue $-1.8 \lesssim \lambda_4 \lesssim -1.4$.  Therefore, the SF-C
transition with $\sigma'=0$ should be XY-like, i.e. $\tilde{\nu} \approx .67$,
$\tilde{\eta} \approx 0.040$, and $\tilde{z} \approx 2$.  Note, however, that
$v'$ is in fact a {\sl dangerous} irrelevant operator, because it is necessary
to select the direction of $\bbox{\psi}$ in the ordered
phase\cite{dangerous_irrelevant_note}.

This asymptotic restoration of the $U(1)$ symmetry should lead to the emergence
of a ``pseudo-Goldstone'' mode near the critical point in the ordered phase.
To see this, consider following the RG flows (for $t'<0$) until the
renormalized reduced temperature is of order one.  At this point, the rescaled
$\bbox{\psi}$ will have amplitude of order one (enforced by the negative
$|\psi|^2$ and positive $|\psi|^4$ terms), but almost unconstrained angle
$\Theta \equiv \tan^{-1}(\psi_2/\psi_1)$.  The renormalized action for $\Theta$
is then of the form
\begin{eqnarray}
    S_{{\rm R},\Theta} & = & \tilde{\xi}^{-3}\int \! d^2\bbox{x}d\tau \bigg\{ 
             \frac{1}{2}\tilde{\xi}^2 \left[\left( \frac{1}{{c'}}
             \partial_\tau \Theta \right)^2 +
             \left( \bbox{\nabla} \Theta \right)^2\right] \nonumber \\
             &&
             -v' \tilde{\xi}^{\lambda_4} \left[\cos^4\Theta +
             \sin^4\Theta\right ]
             \bigg\}. 
\label{theta_action}
\end{eqnarray}
Expanding the cosine and sine terms in $\Theta$ gives the dispersion $\omega^2
= p^2{c'}^2 + \Delta_\perp^2$, with the anomalously small gap $\Delta_\perp^2 =
4v' \tilde{\xi}^{-(2+|\lambda_4|)}$.  Note that the longitudinal gap
$\Delta_\parallel \propto \tilde{\xi}^{-1}$ is much larger than $\Delta_\perp$.
The system appears isotropic in the sense that $\Delta_\perp/\Delta_\parallel
\rightarrow 0$ as $\tilde{\xi} \rightarrow \infty$. A detailed study of
the effect of cubic anisotropy on the critical behavior below $T_c$ can be
found in Ref.~\cite{TS93}.

Until now we have neglected the interaction with the phase $\theta$.
Remarkably, the XY fixed point is stable against these fluctuations.  As
described in detail in Ref.\onlinecite{BN95}, the relevance of $\sigma^\prime$
at the ``classical'' critical point is determined by the specific heat exponent
$\tilde{\alpha} = 2 - D\tilde{\nu}$.  In particular, the eigenvalue
$\lambda_\sigma = \tilde{\alpha}/(2\tilde{\nu})$ is slightly negative at the
three dimensional XY fixed point, and the long wavelength physics at the
critical point is just that of a decoupled $D=3$ XY critical point and a free
massless boson (phase) field.

Another way to understand this result is to integrate out the density
fluctuations $\bbox{\psi}$ and determine their effect upon the superfluid
modes.  This is easily accomplished perturbatively in $\sigma'$, since the
coupling is precisely to the energy operator $|\psi|^2$.  At order $\sigma'$,
one produces only a boundary ($\int i\partial_\tau \theta$) contribution.  The
first non-trivial term occurs at $O({\sigma'}^2)$, where one finds a contribution
to the effective action
\begin{eqnarray}
    \Delta S_1 & = & \frac{{\sigma'}^2}{2}
    \int_{1,2} \partial_\tau\theta_1\partial_\tau\theta_2 \left\langle
    |\psi|^2_1 |\psi|^2_2 \right\rangle_C \nonumber \\
    & \sim &
    {\sigma'}^2 \tilde{\xi}^{2(1-\tilde{\alpha})/\tilde{\nu}} \int_{1,2} 
    \partial_\tau\theta_1\partial_\tau\theta_2
    {\cal C}[|\bbox{x}_1-\bbox{x}_2|/\tilde{\xi},
    |\tau_1-\tau_2|/\tilde{\xi}] \nonumber \\
    & \sim &
    {\sigma'}^2 \tilde{\xi}^{\tilde{\alpha}/\tilde{\nu}}
    \int_{\bbox{x},\tau} |\partial_\tau\theta|^2,
\label{velocity_correction}
\end{eqnarray}
where the subscripts $1,2$ are introduced as a shorthand notation for the
integration variables $\bbox{x}_1,\tau_1$ and $\bbox{x}_2,\tau_2$, ${\cal C}$
is a (short-range) scaling function describing the energy--energy correlations,
and the last line is obtained via a gradient expansion of $\theta_2$.  The
final correction term may be interpreted as a renormalization of the second
sound velocity,
\begin{equation}
{{v'}^2_{\rm R}} = 
\frac{{v'}^2}{{1 + C{\sigma'}^2 m^2 \xi^{{\tilde \alpha}/{\tilde \nu}}
{v'}^2/\tilde{\rho}_s}},
\label{velocity_renormalization}
\end{equation}
where $C$ is an order one constant.  Note that for ${\tilde \alpha}<0$, as is
the case here, this is a small reduction to the velocity near the critical
point.  For ${\tilde \alpha}>0$, however, the correction diverges, and the $v'$
appears to be driven to zero!  This will be the case for the SF-X transition
discussed in the next section and requires a more careful analysis.  Here we
simply point out that the second sound velocity undergoes a finite {\sl
  suppression} by critical fluctuations (note that, once ${\tilde \alpha}$ is
negative, non-singular contributions actually dominate over the singular one
exhibited explicitly in Eq.(\ref{velocity_renormalization}), and lead to a
finite correction to $v'_{\rm R}$).


\section{Superfluid--Checkerboard Supersolid Transition}
\label{SF_X_transition}

In this section we discuss the situation when $t' \gg 1, u'$, but $t$ passes
through zero.  In this case it is the $\bbox{\psi}$ modes which are weakly
fluctuating, and the resulting theory is simply that of the $\Phi$ field
coupled to the superfluid phase.  Here the critical behavior for $\sigma=0$ is
simply that of a classical three-dimensional Ising model.  However, because the
specific heat exponent in this case is positive ($\alpha \approx 0.12$), these
phase fluctuations are {\sl relevant}, and the problem needs to be
reconsidered.

To do so, we employ the technique of $\varepsilon$--expansion, developing RG
equations near the upper critical dimension $D_{\rm uc}=d_{\rm uc}+1=4$.  In
this section, we proceed with the simplest such method, known as a
momentum-shell RG, which is sufficient to extract most aspects of the critical
behavior.  In section \ref{dim_reg}, we use more powerful field theoretic
methods within dimensional regularization to compute full crossover quantities.
Here the theory is regularized by including a sharp cut-off in momenta: all the
functional integrals in the partition function $Z$ are only over fields with
momenta $|q|<\Lambda$, the cut-off wavevector.  Such a cut-off serves as a
crude approximation to the physical lattice cut-off of the bosonic system, but
the precise form is irrelevant for the perturbative RG to the order required
here, and facilitates the calculation of certain quantities.

The momentum shell RG is performed by successive elimination of shells of
momenta near the cut-off $\Lambda$.  After each shell is integrated out of the
partition function, the coordinates and momenta are rescaled to keep the value
of the cut-off fixed.  Additional rescalings of the fields allows most of the
quadratic part of the action to remain fixed as well.  The remaining
coefficients {\sl flow} under the RG.  While an infinite number of operators
are generated, only the relevant (and marginal) ones need be kept within the
$\varepsilon$ expansion, in this case $u$, $\sigma$, and $v$.

Momentum shell RG suffers, unfortunately, from one flaw: it is unable to deal
properly with situations in which loop integrals must be handled at non--zero
external momenta (and frequencies).  For most problems, such momentum and
frequency dependence can indeed be neglected to $O(\varepsilon)$ in the RG.  In
a cubic theory such as this one, however, the second diagram in
Fig.~\ref{gamma_02}\ can potentially produce corrections at $O(\varepsilon)$,
because the coupling $\sigma$ is only $O(\varepsilon^{1/2})$.  We investigate
this diagram in appendix \ref{loop_appendix}, and find that these dangerous
contributions are actually not present.  The momentum shell RG is therefore
well defined to leading order in $\varepsilon$, and we proceed here with the
remainder of the calculation.

We define the mode elimination step of the RG by integrating out fields with
momenta in the range $\Lambda/b < p < \Lambda$, and use the critical dynamics
method of integrating out {\sl all} frequencies, which are not cut off.  By
choosing $b=e^{-dl}$, where $dl$ is infinitesimal, we arrive at differential RG
flow equations for the coupling constants.  These equations are easily computed
diagrammatically.  Because we evaluate all loops at zero external momenta (c.f.
the previous paragraph), the constraint that {\sl both} loop momenta be within
the shell is trivially satisfied by constraining just the single integration
variable, due to momentum conservation.

Consider first the diagram in Fig.~\ref{gamma_20}.  Because of the explicit
imaginary time derivative in the coupling $\sigma$, this diagram generates a
renormalization of the velocity $v$.  Because of the explicit factor of $i$ in
the vertex, it decreases $v$ as in the perturbative treatment of the previous
section.  Because this is the {\sl only} renormalization of $v$ at one loop,
the search for possible critical fixed points must proceed {\sl in the limit $v
  \rightarrow 0$}.  This simplifies the computation of certain diagrams.  In
particular, any internal $\theta$ lines may be treated in this limit as simply
constant $v^2$ factors.  All remaining loop integrals are then proportional to
the standard form
\begin{equation}
\int_{\Lambda/b<p<\Lambda} 
\frac{{d^3\bbox{p}}}
     {{(2\pi)^3}} \int_{-\infty}^\infty \frac{{d\omega}}{{2\pi}} 
     \frac{1}{{(p^2+\omega^2/c^2)^2}} = A_3 c dl/4,
\label{simple_loop_integral}
\end{equation}
where $A_3 = 1/(2\pi^2)$ is a geometric factor.  To complete the calculation,
the fields and coordinates are rescaled to keep the cut-off fixed, according to
\begin{eqnarray}
  x \rightarrow b x \, , & \hskip 0.5truein & \tau  \rightarrow  b^z
  \tau \, ,\\
  p  \rightarrow  p/b \, ,& & \omega \rightarrow  \omega/b^z \, ,\\
  \Phi  \rightarrow  \Phi b^{(2-d-z-\eta)/2} \, ,& & \theta  \rightarrow
  \theta b^{(2-d-z-\tilde{\eta})/2} \, ,
\label{rescalings}
\end{eqnarray}
where in this case, because of the absence of momentum-dependent loop
corrections, $\eta = \tilde{\eta} = z-2 = 0$ (to this order).  The flow
equations resulting from the rescaling and mode elimination are
\begin{eqnarray}
  \frac{{dK}}{{d\ell}} & = & (1 - U/4 - 5 K / 2) K \label{Kflow} \, ,\\
  \frac{{dU}}{{d\ell}} & = & U - (3U^2/8 +
  6 KU + 24 K^2) \label{Uflow} \, , \\
  \frac{{d\tilde{V}}}{{d\ell}} & = & -  K V/4 \label{vflow} \, ,\\
  \frac{{d(t-t_0)}}{{dl}} & = & [2 - \varepsilon(U/8+K)](t-t_0) \, , 
\label{tflow}
\end{eqnarray}
where we have defined $\varepsilon = 3-d$, $\ell = \varepsilon l$, and the
dimensionless couplings $K = A_3 \sigma^2 c v^2/\varepsilon$, $U = A_3 u
c/\varepsilon$, and $V = v/c$.  The $T_c$ shift $t_0 = - \varepsilon
\Lambda^2(U/8+K)$.  The RG calculation is controlled for $d=3-\varepsilon$,
where the coupling constants are of order $\varepsilon$, and higher loop
corrections to the flow equations are higher order in $\varepsilon$.  Note that
$K$ is strictly positive, so $V$ indeed flows monotonically to zero.

The flow diagram resulting from Eqs.~(\ref{Kflow})--(\ref{Uflow})\ is shown in
Fig.~\ref{mom_shell_flow_fig}.  The three-dimensional classical Ising fixed
point ($U=8/3$,$K=0$) is unstable once phase fluctuations are introduced.
However, a new weak-coupling fixed point is not found within a one-loop
$\varepsilon$--expansion. Instead, the couplings flow into the classically
unstable regime where $U<0$.  This scenario is generally interpreted as a
fluctuation-induced first order phase transition \cite{Ami84}. As was mentioned
in the introduction, the existence of a strong coupling fixed point can not be
excluded within the above line of argument. In fact, upon using a field
theoretic RG approach carried out in fixed dimension, we will show in the next
section that a strong coupling fixed point can be identified for large enough
values of the coupling constant $V$. The resulting ``non-Bose liquid''
behavior will be discussed in the next section.

For small values of $V$ ``runaway trajectories'' persist. In that case the
instabilities in the trajectories may indicate a fluctuation-driven first order
transition may actually describe the physics of the supersolid transition. One
can understand this by matching the RG flows onto a renormalized classical
theory.  Close enough to the critical point (the flows may be integrated down
to the scale of the correlation length), the renormalized theory has a negative
value of $u$. Mean--field theory is justified for the renormalized action,
since the effective $t$ is now order one.  With a negative $u$, a standard
mean--field analysis then predicts a first order transition before the critical
point is actually reached.  Note that, if the initial coupling $\sigma$ (and
hence $K$) is very small, this will require an extremely close approach to the
critical point, and hence result in extremely small discontinuities in
thermodynamic parameters (such as $\langle\Phi\rangle$).  Experiments or
numerical simulations which do not probe sufficiently close to $t_0$ may not
resolve the true first-order transition, and instead observe ``effective''
exponents (or more complex cross-over behavior).  At a true fixed point, the
correlation length exponent would be determined from Eq.~(\ref{tflow})\ as
$1/\nu = 2 - \varepsilon(U/8+K)$.  A rough estimate of the measured value of
$\nu$ may be obtained from an examination of Fig.~\ref{mom_shell_flow_fig}.
Using values of $U$ and $K$ along a flow from the (unstable) Ising fixed along
the central trajectory in the figure suggests a gradual reduction of $\nu_{\rm
  eff.}$ from approximately $0.6$ ($\nu = 0.6$ is the $O(\varepsilon)$ result
for the Ising value -- the accepted value is $\nu \approx 0.64$) towards $0.5$.


\section{Perturbation Theory and Renormalization: Dimensional Regularization}
\label{dim_reg}

In this section we use methods adapted from field theory to
investigate the critical behavior of the superfluid to checkerboard
supersolid transition. This allows us to go beyond the results
obtained in section \ref{SF_X_transition} and study the full crossover
behavior near criticality. Furthermore, we can study the flow of the
parameters and coupling constants for arbitrary values of the second
sound velocity. This will turn out to be an important technical
advantage, since it allows us to identify a strong coupling fixed point
above a critical value of $v/c$ (see below).

A perturbation theory for the effective free energy functional of the
supersolid phase may be set up following the common procedure (see e.g.
Ref.~\cite{Ami84}). The elements of perturbation theory are the free
propagators for the one-component density order parameter $\Phi$
\begin{equation}
  G_{\Phi \Phi} ({\bf k}, \omega) =
  \frac{1}{t_0 + (\omega/c_0)^2 + k^2} \, ,
\end{equation}
and the superfluid phase
\begin{equation}
  G_{\Theta \Theta} ({\bf k}, \omega) =
  \frac{1}{(\omega/v_0)^2 + k^2}\, ,
\end{equation}
and the vertices, which may be read off from the anharmonic parts of
Eqs.~(\ref{critical_action})-(\ref{ptcoupling}).  In
Fig.~\ref{elements_perturbation_theory} we depict these elements for
constructing the Feynman graphs of the field-theoretical representation of the
critical behavior of a supersolid. Following the standard notation in field
theory bare values are indicated by a subscript ``0''.

\subsection{Renormalization factors}

For $t_0 \rightarrow 0$ the perturbation theory is infrared-divergent
leading to non-trivial critical exponents. These anomalous dimensions are
derived by studying the ultraviolet singularities of the field theory, which
appear at the upper critical dimension $d_c=3$, when the momentum cutoff
$\Lambda$ is pushed to infinity (see e.g. Ref.\cite{Ami84}). Finite values are
then assigned to these UV-divergent integrals through the application of a
regularization prescription. We shall choose the dimensional regularization
scheme as introduced by t'Hooft and Veltman \cite{Hoo72}; here the ($\Lambda
\rightarrow \infty$) singularities appear as poles $\propto 1 / (d_c-d)$.

The ultraviolet divergences may then be collected in renormalization constants 
and absorbed into the definition of multiplicatively renormalized quantities. 
Thus we define the renormalized fields
\begin{equation}
              \Phi   = Z_\Phi^{1/2}         \Phi_0 \quad , \qquad  
              \Theta = Z_{\Theta}^{1/2}     \Theta_0 \quad , 
\end{equation}  
and the renormalized parameters
\begin{eqnarray}
     t   &&= Z_\Phi^{-1}   Z_t (t_0 - t_{0c}) \mu^{-2} \, , \label{Z_t}\\
     c^2 &&= Z_\Phi        Z_c^{-1}     c_0^2          \, , \label{Z_c}\\     
     v^2 &&= Z_\Phi        Z_v^{-1}     v_0^2          \, , \label{Z_v}\\
     u   &&= Z_\Phi^{-2} Z_u u_0 \mu^{d-3} S_d         \, , \label{Z_u}\\
  \sigma &&= Z_{\Theta}^{-1/2} Z_\Phi^{-1} Z_\sigma \sigma_0 
             \mu^{(d-3)/2} S_d^{1/2}                   \, . \label{Z_sigma}
\end{eqnarray}
In Eq.~(\ref{Z_t}) we have taken into account the fact that the fluctuations
will also shift the transition temperature. Furthermore we have rendered the
renormalized quantities dimensionless by introducing the explicit arbitrary
length scale $1/\mu$, and have finally included the geometric factor
\begin{equation}
  S_d = \frac{2}{(4 \pi)^{d/2} \Gamma(d/2)} \, .
\end{equation}
Adapting a {\em fixed dimension RG approach}, one finds from the one--loop
expressions for the singular parts of the vertex functions $\Gamma_{02}$,
$\Gamma_{20}$, $\Gamma_{12}$, and $\Gamma_{04}$ in appendix~\ref{app_pert} the
following results for the renormalization constants:
\begin{eqnarray}
  Z_t = &&1 - \frac{uc}{8 \varepsilon} - \frac{(v \sigma)^2 c}{\varepsilon} 
        \frac{1}{(1 + v/c)^2} \, , \\
  Z_\Phi = &&1 -  \frac{(v \sigma)^2 c}{\varepsilon} 
                  \frac{v/c}{(1+v/c)^3} {\cal A} (v/c,d) \, , \\
  Z_c = &&1 + \frac{2 (v \sigma)^2 c}{\varepsilon} 
          \frac{v/c}{(1 + v/c)^3} \, , \\  
  Z_v = &&1 + \frac{(v \sigma)^2 c}{2 \varepsilon} \, , \\
  Z_u = &&1 - \frac{3 u c}{8 \varepsilon} - 
              \frac{6 (v \sigma)^2 c}{\varepsilon} 
              \frac{1}{(1+ v/c)^2} \nonumber \\
        &&-\frac{24 (v \sigma)^4 c^2}{\varepsilon u c} 
              \frac{1}{(1+ v/c)^3} \, , \\
  Z_\sigma = &&1 - \frac{u c}{4 \varepsilon} 
            - \frac{2 (v \sigma)^2 c}{\varepsilon} \frac{1}{(1+v/c)^2} \, .
\end{eqnarray} 
For the fixed-dimension RG, the quantity ${\cal A}$ depends on both
$V$ and $d$ (see Eq.~(\ref{aA}) in appendix \ref{app_pert}).  As a
consequence of the momentum dependence of the three--point vertex,
there are no singular contributions to the $q^2$--term in the harmonic
part of the superfluid phase field to any order in perturbation
theory. This implies the exact identity
\begin{equation}
  Z_{\Theta} = 1 \, .
\label{Z_exact}
\end{equation}

\subsection{Renormalization group equations and flow functions}

The renormalization group equation serves to connect the asymptotic theory,
where the infrared singularities manifest themselves, with a region in
parameter space where the couplings $u$ and $\sigma$ are finite (but not
necessarily small), and an ordinary ``naive'' perturbation expansion becomes
applicable. It explicitly takes advantage of the scale invariance of the system
near the critical point. More precisely, we observe that the bare two-point
vertex function is of course independent of the arbitrary renormalization scale
$\mu$:
\begin{equation}
      \mu \frac{d}{d \mu} \bigg \vert_0
      \Gamma^0_{ln}(r_0,c_0,v_0,u_0,\sigma_0,{\bf q},\omega) = 0 \, . 
\label{4.1}
\end{equation}
Introducing Wilson's flow functions
\begin{eqnarray}
     \zeta_\Phi &&=  {\cal D}_\mu \ln Z_\Phi \, , \quad \quad 
     \zeta_{\Theta} =  {\cal D}_\mu \ln Z_{\Theta}                          
     \, , \label{4.2}\\
     \zeta_t &&= {\cal D}_\mu \ln \frac{t}{t_0 - t_{0c}} 
               = - 2 - \zeta_\Phi + {\cal D}_\mu \ln Z_t    \, , \label{4.3}\\
     \zeta_c &&= {\cal D}_\mu \ln \frac{c}{c_0} 
               = \frac{1}{2} \zeta_\Phi - \frac{1}{2} {\cal D}_\mu \ln Z_c    
                                                            \, , \label{4.4}\\
     \zeta_v &&= {\cal D}_\mu \ln \frac{v}{v_0} 
               = \frac{1}{2} \zeta_\Theta 
                 - \frac{1}{2} {\cal D}_\mu \ln Z_v      \, , \label{4.5}\\
     \zeta_\sigma  &&= {\cal D}_\mu \ln \frac{\sigma}{\sigma_0} 
               = \frac{d-3}{2} 
                 - \frac{\zeta_{\Theta}}{2} 
                 - \zeta_\Phi   
                 +  {\cal D}_\mu \ln Z_\sigma            \, , \label{4.6}\\
     \zeta_u &&= {\cal D}_\mu
                 \ln \frac{u}{u_0} = (d - 3)
                 -  2 \zeta_\Phi 
                 + {\cal D}_\mu \ln Z_u                   \, , \label{4.7}
\end{eqnarray}
where ${\cal D}_\mu = \mu \frac{\partial}{\partial \mu} \vert_0$ denotes a
logarithmic derivative with fixed bare couplings and parameters,
Eq.~(\ref{4.1}) leads to renormalization group equations for the renormalized
vertex functions $\Gamma_{ln}$. For instance, the two--point
vertex function of the Ising fields $\Phi$ satisfies the partial differential equation 
\begin{eqnarray}
     \Biggl[  &&\mu \frac{\partial}{\partial \mu} + 
             \zeta_t t \frac{\partial}{\partial r} + 
             \zeta_c c \frac{\partial}{\partial c} +
             \zeta_v v \frac{\partial}{\partial v} +
             \zeta_\sigma \sigma \frac{\partial}{\partial \sigma} \nonumber\\ +
             &&\zeta_u u \frac{\partial}{\partial u} + \zeta_\Phi \Biggr] 
                              \Gamma_{02}(\mu,t,c,v,u,\sigma,{\bf k},\omega) 
             = 0 \, .
\label{4.8}
\end{eqnarray}

The RG equation (\ref{4.8}) is now readily solved with the method of
characteristics. The characteristics $a(s)$ of Eq.~(\ref{4.8}) define the
running parameters and coupling constants into which these transform when $\mu
\rightarrow \mu(s) = \mu s$. They are given by the solutions to
first-order differential equations ($a=t,c,v,u,\sigma$)
\begin{equation}
  s \frac{d a(s)}{d s} = \zeta_a(s) a(s) \, ,
  \label{4.9}
\end{equation}
with the initial conditions $t(1) = t$, $c(1) = c$, $v(1) = v$, $u(1) = u$, and
$\sigma(1) = \sigma$, namely
\begin{equation}
     a(s) = a \exp \left[ \int_1^s \zeta_a(s^\prime) d s'/ s' \right] \, . 
\label{4.10}
\end{equation}
Defining the dimensionless vertex function ${\hat \Gamma}_{02}$ according to
\begin{equation}
     \Gamma_{02}(\mu,t,c,v,u,\sigma,{\bf k},\omega) = 
       \mu^2 {\hat \Gamma}_{02} 
       \left(   t,u,\sigma, \frac{v}{c}, \frac{{\bf k}}{\mu},
                \frac{\omega^2}{c^2 \mu^2}
       \right) \, , 
\label{4.11}
\end{equation}
the solution of Eq.~(\ref{4.8}) is
\begin{eqnarray}
      &&\Gamma_{02}(\mu,t,c,v,u,\sigma,{\bf k},\omega) = 
      \mu^2 s^2 \exp \left[ \int_1^s \zeta_\Phi(s') d s' / s' \right]
      \nonumber \\ &&\times
      {\hat \Gamma_{02}} 
      \left( t(s), u(s), \sigma(s), \frac{v(s)}{c(s)}, 
             \frac{{\bf k}}{\mu s},
             \frac{\omega^2}{c(s)^2 \mu^2 s^2}
      \right) \, . 
\label{4.12}
\end{eqnarray}

\subsection{Flow diagram and fixed points}
To one-loop order Wilson's flow functions as derived from
Eqs.~(\ref{4.2})-(\ref{4.7}) read
\begin{eqnarray}
  \zeta_\Phi  = && (v \sigma)^2 c \frac{v/c}{(1+v/c)^3} 
                  {\cal A} (v/c,d) \, , \\
  \zeta_c     = &&\frac{1}{2} \zeta_\Phi +  
                  (v \sigma)^2 c \frac{v/c}{(1+v/c)^3} \, , \\  
  \zeta_v     = &&\frac{1}{2} \zeta_\Theta + \frac{1}{4} (v \sigma)^2 c \, , \\
  \zeta_t     = &&-2 - \zeta_\Phi + \frac{1}{8} uc
                  +(v \sigma)^2 c \frac{1}{(1+v/c)^2}   \, , \\
  \zeta_\sigma= &&\frac{d-3}{2} -\frac{\zeta_{\Theta}}{2} - \zeta_\Phi 
                 +\frac{1}{8} uc + (v \sigma)^2 c \frac{1}{(1+v/c)^2} \, , \\
  \zeta_u     = &&(d-3) - 2 \zeta_\Phi + \frac{3}{8} uc  
                  +6 (v \sigma)^2 c \frac{1}{(1+v/c)^2} \nonumber \\
               +&&24 \frac{ (v \sigma)^4 c^2}{u c} \frac{1}{(1+v/c)^3} \, ,
\end{eqnarray}
and due to the exact relation Eq.~(\ref{Z_exact}) one finds
\begin{equation}
  \zeta_\Theta = 0 \, .
\end{equation}
The zeros of the $\beta$--functions,
\begin{eqnarray}
  \beta_u = &&\mu \frac{ \partial}{\partial \mu} \bigg \vert_0 u 
          = u \zeta_u \, ,\\
  \beta_\sigma = &&\mu \frac{\partial}{\partial \mu} \bigg \vert_0 \sigma 
          = \sigma \zeta_\sigma \, ,
\end{eqnarray}
determine the values of the coupling constants where the theory becomes scale
invariant.

In order to study the flow diagram, it is convenient to introduce new effective
coupling constants $K := \sigma^2 v^2 c$, $U := uc$, and $V := v/c$, whose flow
equations can be readily obtained from Eqs.~(\ref{4.2})--(\ref{4.7}).
They are
\begin{eqnarray}
   \frac{1}{K}\frac{d K}{d \ell}
   &=& \varepsilon\!-\!\frac{U}{4}\!-\!
       K \left[\!\frac{2}{(1\!+\!V)^2} + \frac{1}{2} + 
                \frac{V(1\!-\!3{\cal A}/2)}{(1\!+\!V)^3}\!\right], 
   \label{rg1} \\
   \frac{1}{U}\frac{d U}{d \ell}
   &=& \varepsilon - \frac{3}{8}U - 24 \frac{K^2}{U(1+V)^3 } \nonumber\\ 
   &&-K \left[ \frac{6}{(1+V)^2} + \frac{V(1-3{\cal A}/2)}{(1+V)^3}  
         \right], 
   \label{rg2} \\
   \frac{1}{V}\frac{d V}{d \ell}
   &=& - K \left[ \frac{1}{4} - \frac{V(1+{\cal A}/2)}{(1+V)^3} \right],
   \label{rg3}
\end{eqnarray}
where, for the purpose of comparison with Eqs.~(\ref{Kflow}-\ref{vflow}) we
have defined the running {\sl logarithmic} scale $\ell = - \ln s$, so that
$d/d\ell = -sd/ds$. Note that ${\cal A}$ depends on $V$ and $d$ (see
Eq.~(\ref{aA}) in appendix \ref{app_pert}). Depending on the sign of the
expression in the square brackets of Eq.~(\ref{rg3}), ${\cal B}(V,d) = 1/4 -
V(1+{\cal A}(V,d)/2)/(1+V)^3$, the flow of $V$ either tends to $V=0$ or to
$V=\infty$.  One finds that there is a critical value $V_c (d)$, where ${\cal
  B}(V,d)$ changes sign. One of the most essential features of $V_c (d)$ is
that it cannot be obtained within a strict $\varepsilon$--expansion. As can be
inferred from Fig.~\ref{v_crit} there is a critical value $d_c \approx 2.598$
above which there is no solution of the equation ${\cal B}(V,d) = 0$. For
$d<d_c$ there are two real roots, where one diverges upon approaching $d=2$.
Since we are primarily interested in the critical behavior in $d=2$ dimensions
the most relevant root is given by the lower curve in Fig.~\ref{v_crit}. In
what follows we will restrict our discussion to the case $d=2$. Then the
flow--equations have the following fixed points
\begin{equation}
\begin{array}{llll}
  V = 0  \, , \,   &K = 0 \, ,  \,  &U = 0   
  \, &{\rm (Gaussian \, I)} \, ,  \\
  V = 0  \, ,  &K = 0 \, , \,\, &U = 8 / 3   
  \,\,&{\rm (Ising \, I)} \, . \\
  V = \infty  \, , \,\, &K = 0 \, ,\,\, &U = 0   
  \,\,&{\rm (Gaussian \, II)} \, , \\
  V = \infty  \, , \,\, &K = 0 \, ,\,\, &U = 8/3  
  \,\,&{\rm (Ising \,II)} \, , \\
  V = \infty  \, , \,\, &K = 4 / 3  \, ,\,\, 
  &U = 16 / 3  
  \,\,&{\rm (Non\!\!-\!\!Bose \, liquid)} \, .
\end{array}
\end{equation}
The crucial quantity for the behavior of the coupling constants is the ratio,
$V = v/c$, of the velocity of the second sound $v$ and phonon velocity $c$
(first sound). For $V<V_c^1(d=2) \equiv V_c$ it turns out that the flow of $V$
tends to zero, in which case {\sl both} the Gaussian I and the Ising I fixed
points are unstable to small perturbations. This corresponds to the runaway
trajectories discussed in the preceding section. However, the behavior of the
renormalization group trajectories become quite different for $V>V_c$. Then the
flow of $V$ tends to infinity, but it still depends on the sign of the
coefficients in the flow equations for $U$ and $K$, whether the renormalization
group trajectories show runaway flow or converge into a finite fixed point.
In the limit $V\rightarrow \infty$ the flow equations reduce to the following
simple form
\begin{eqnarray}
   \frac{1}{K}\frac{d K}{d \ell}
   &=& 1 - \frac{1}{4}U + \frac{1}{4}K \, ,
   \label{rginf1} \\
   \frac{1}{U}\frac{d U}{d \ell}
   &=& 1 - \frac{3}{8}U + \frac{3}{4}K \, ,
   \label{rginv2} 
\end{eqnarray}
the flow diagram of which is shown in Fig.~\ref{flow_non_bose_liquid}.

The manifold, which separates the region in parameter space, where the flow of
the renormalization group trajectories ends up in the non-Bose liquid fixed
point (stable trajectory), from the parameter space where the flow shows runaway
behavior (unstable trajectory), is called the separatrix. A visualization of
the complex flow--behavior is given in Fig.~\ref{figure5}, where we have
chosen the renormalization group trajectories to start at the plane $V=3$
(bottom of the box). This part of the parameter space shows stable as well as
unstable trajectories, where all stable trajectories are attracted to a
critical surface, which finally contracts to the non-Bose liquid fixed point. A
numerical analysis of the flow equations also shows that approximately below
$V=2$ there are no more stable trajectories. In Figs.~\ref{figure6} and
\ref{figure7} two--dimensional projections of the flow diagram are shown with
starting values $V=3$ and $V=1$, respectively.

\subsection{Scaling behavior and effective exponents}

Now we turn to the investigation of Eq.~(\ref{4.12}) near a fixed point ${\bf
  G}^\star = (K^\star, U^\star, V^\star)$; introducing the fixed--point values
of the Wilson flow functions $\zeta_a^\star = \zeta_a ({\bf G}^\star)$, also
called the anomalous dimension of the parameter $a$, one finds the following
scaling law
\begin{eqnarray}
      &&\Gamma_{02}(\mu,t,{\bf G},{\bf k},\omega) = 
      \mu^2 s^{2 + \zeta_\Phi^\star}
      \nonumber \\ &&\quad \times
      {\hat \Gamma_{02}} 
      \left( t s^{\zeta_t^\star}, {\bf G}^\star, \frac{{\bf k}}{\mu s},
            \frac{\omega^2}{c^2 \mu^2 s^{2 + 2 \zeta_c^\star}} 
      \right) \, . 
\end{eqnarray}
For $\Gamma_{20}$ one obtains a similar scaling law, where $\zeta_\Phi$ and
$\zeta_c$ are replaced by $\zeta_\Theta$ and $\zeta_v$, respectively. Using the
matching condition $k / \mu s = 1$ the following dynamic scaling law,
\begin{eqnarray}
      &&\Gamma_{02}(\mu,t,{\bf G},{\bf k},\omega) \propto 
      q^{2 - \eta_\Phi}
      \nonumber \\ &&\quad\times
      {\hat \Gamma_{02}} 
      \left( \frac{t}{(k/\mu)^{1/\nu}}, {\bf G}^\star, 1,
             \frac{\omega^2}{\mu^2 c^2 (k/\mu)^{2z}} 
      \right) \, , 
\end{eqnarray}
is obtained, where the {\sl five} independent exponents
\begin{eqnarray}
  \eta_\Theta &=& - \zeta_\Theta^\star = 0 \, ,\\ 
  z_\Theta    &=&  1 + \zeta_v^\star \, ,\\
  \eta_\Phi   &=& - \zeta_\Phi^\star \, ,\\ 
  z_\Phi      &=&  1 + \zeta_c^\star \, ,\\
  \nu  &=& -\frac{1}{\zeta_t^\star} \, , \\
  \Delta &=& \zeta_v^\star - \zeta_c^\star \, . \\
\end{eqnarray}

All other exponents are found through the scaling relations
$\gamma = \nu (2 - \eta)$ and $2 \beta + \gamma = 2 - \alpha = \beta
(1+\delta)$.

The crossover exponent $\Delta$ describes the flow of the ratio of the two
sound velocities, $V(s) = v(s) / c(s) \rightarrow s^\Delta v/c$. As discussed
in the previous section there is a separatrix in parameter space which
separates stable from unstable trajectories. In the unstable region all
trajectories show runaway behavior, but depending on the initial value of $V$
being smaller or larger than $V_c$ the flow of $V$ either tends to zero or to
infinity.

Let us consider two extreme situations: For $V = 0$ we find $\beta_K = - K +
\frac{5}{2} K^2 + \frac{1}{4}KU$ and $\beta_U = - U + \frac{3}{8} U^2 + 6 K U +
24 K^2$ with the corresponding fixed points:
\begin{equation}
\begin{array}{lll}
  K = 0 \, , \quad &U = 0   
  \quad&{\rm (Gaussian \, I)} \, ,  \\
  K = 0 \, , \quad &U = \frac{8}{3}
  \quad&{\rm (Ising \, I)} \, .
\end{array}
\end{equation}
The fixed point values for the flow--functions at the Ising fixed point are
$\zeta_\Phi^\star = 0$, $\zeta_c^\star = 0$, $\zeta_v^\star = 0$, and
$\zeta_t^\star = -2 + \frac{1}{3}$. Hence we find the following
critical exponents at the Ising fixed point (to one--loop order):
\begin{eqnarray}
   \eta_\Phi &=& 0   \, , \\ 
   z_\Theta = z_\Phi   &=& 1  \, , \\
   \nu  &=& \frac{1}{2} + \frac{1}{6} \, .
\end{eqnarray}
For $V = \infty$ we find in $d=2$ dimensions $\beta_K = - K - \frac{1}{4} K^2 +
\frac{1}{4}KU$ and $\beta_U = - U + \frac{3}{8} U^2 - \frac{3}{4}KU$ with the
corresponding fixed points:
\begin{equation}
\begin{array}{lll}
  K = 0, \,  &U = 0   
  \, &{\rm (Gaussian \,  II)}, \\
  K = 0, \, &U = 8 / 3  
  \, &{\rm (Ising \, II)}, \\
  K = 4 / 3,  \, &U = 16 / 3  
  \, &{\rm (Non\!-\!Bose \, liquid)}.
\end{array}
\end{equation}
The fixed point values of the flow--functions at the non-Bose liquid
fixed point read $\zeta_\Phi^\star = 2/3$, $\zeta_c^\star = 1/3$,
$\zeta_v^\star = 1/3$, $\zeta_t^\star = -2$ and $\zeta_\sigma = 1/2$,
where interestingly all fluctuation contributions to the
renormalization of $t$ cancel. Hence we find the following critical
exponents at the non-Bose liquid fixed point (to one--loop order):
\begin{eqnarray}
   \eta_\Phi &=& - \frac{2}{3}  \, ,\\ 
   z_\Phi  &=& \frac{4}{3} \, ,\\
   z_\Theta  &=& \frac{4}{3} \, , \\
   \nu  &=& \frac{1}{2} \, .
\end{eqnarray}

Now, we consider crossover effects on the most interesting physical quantities,
namely the correlation functions of the density order parameter and the
superfluid phase. For the discussion we will use an approximation, frequently
called renormalized mean field theory, where one uses the fully renormalized
flow--functions, but the zero--loop result for the scaling function. Thus one
gets 
\begin{eqnarray}
  \Gamma_{\Theta \Theta} = \Gamma_{20} (t, {\bf q} ) 
  &=& \mu^2 s^2 \exp \left[{\int_1^s \zeta_{\Theta} (s^\prime) d s^\prime /
  s^\prime } \right] \nonumber \\
  &&\quad \times \left[  \frac{k^2}{\mu^2 s^2} + 
                         \frac{\omega^2}{\mu^2 s^2 v^2(s)} 
                 \right] \, , \\
  \Gamma_{\Phi \Phi} = \Gamma_{02} (t, {\bf q} ) 
  &=& \mu^2 s^2 \exp \left[ \int_1^s \zeta_{\Phi} (s^\prime) d s^\prime /
  s^\prime \right] \nonumber \\
  &&\quad \times \left[ t (s) + \frac{k^2}{\mu^2 s^2} + 
                        \frac{\omega^2}{\mu^2 s^2 c^2(s)} 
                \right] \, .
\end{eqnarray}
The most convenient way to analyze the crossover behavior of the correlation
functions is in terms of effective critical exponents. We consider first the
case $t=0$ and choose to define the effective dynamic critical exponents
\begin{eqnarray}
  z_{\Theta} (s) &=& 1 + \zeta_v (s) \, , \\
  z_{\Phi}   (s) &=& 1 + \zeta_c (s) \, , 
\end{eqnarray}
which are shown in Fig.~\ref{figure8} as a function of the flow parameter $s$,
where for the initial values we have chosen $U(s = 1) = 8/3$, $K(s = 1) =
1/100$ and a series of ratios of sound velocities $V(1) = 10,5,3$. Upon using
the matching condition $q/\mu s = 1$ the flow parameter $s$ can be related to
the wave number. A corresponding effective Fisher exponent for the Ising field
can also be defined by $\eta_\Phi = - \zeta_\Phi (s)$. The crossover behavior
is shown in Fig.~\ref{figure9} for the same set of initial values as for the
dynamic critical exponents.

At zero external frequencies and momenta one can define an effective critical
exponent for the dependence of the Ising field correlation function on the
control parameter
\begin{eqnarray}
  v(s) = - 1 / \zeta_t (s) \, .
\end{eqnarray}
Using again the same set of initial values for the coupling constants one
obtains the crossover behavior depicted in Fig.~\ref{figure10}.

\section{Summary and Conclusions}

We have studied the SS-SF critical phenomena using renormalization group
techniques.  Physical results are obtained by extrapolating from $d=3-\epsilon$
to $d=2$, corresponding to bosons on a periodic substrate.  Coupling to the
gapless phase modes of the superfluid opens the possibility of unusual
quantum-critical behavior.  This is {\sl not} realized for the CSS-SF
transition, in which the coupling of critical and phase modes is {\sl
  irrelevant}, and the universality class is simply three-dimensional XY-like.
By contrast, the mode-coupling terms are {\sl relevant} for the XSS-SF
transition, and generically modify the critical behavior.  The $\epsilon$
expansion predicts a fluctuation-driven first order transition in this case.  A
fixed dimension RG, however, gives a different scenario. For small values of
the ratio $V = v/c$ runaway trajectories persist indicating again a first order
transition though we cannot rule out the possibility of a strong-coupling
continuous critical point. For large values of $V$ we have identified a strong
coupling fixed point. If the latter occurs, the phase and critical density
modes must be strongly coupled at the transition, resulting in anomalous
superfluid behavior. At the non-Bose liquid fixed point the exponent for the
correlation length is found to be $\nu = 1/2$ to one loop order, since all
fluctuation corrections cancel. It would be interesting to check whether there
are two-loop contributions to this classical value. The value for the Fisher
exponent $\eta_\Phi = -2/3$ is negative, similar as in the case of a normal to
superconducting transition~\cite{NSzf}.  In either case, there are special
multi-critical points in the phase diagram (at particular values of the
chemical potential) for which the mode-coupling terms explicitly vanish by
particle-hole symmetry, and simple 3d Ising behavior (and ordinary
superfluidity) is restored.

An attempt to extract critical properties at the XSS-SF transition in the
quantum phase model has been made in Ref.~\onlinecite{vOW95}.  In this model
the multi-critical point may be exactly located at $n_0 = 1/2$ using
particle-hole symmetry.  There, they find a correlation exponent $\nu = 0.55
\pm 0.05$.  This is somewhat less than the expected 3d Ising behavior $\nu
\approx 0.64$.  They also quote a value $z=1$ here, in agreement with the RG.
We believe the disagreement with the 3d Ising result at the particle-hole
symmetric point (where the theory appears to be on extremely firm ground) is
probably due to finite-size corrections in the numerical data.  Away from this
point (they study $n_0 = 0.4$), the correlation length exponent is perhaps
reduced, $\nu = 0.5 \pm 0.11$, although the larger error bars make this
interpretation difficult.  A reduction of $\nu$ from $\nu \approx 0.64
\rightarrow 0.5$ is indeed predicted by the crossover analysis for a scenario,
where a non-Bose liquid fixed point exists. In the latter case the
dynamic critical exponent is found to be $z=4/3$ to one-loop
order. Interestingly, scaling analysis of their Monte-Carlo data does
not appear to differentiate between a dynamical scaling exponent $z=1$
or $z=2$.  However, they put a stronger constraint on the combination
$z+\eta = 0.8 \pm 0.2$, which is consistent with $z+\eta = 4/3-2/3
\approx 0.67$.  This gives some support to the RG prediction, though
further systematic analysis of the simulation data is certainly
desirable.

More recently, Scalettar {et\ al.}\cite{SZ96}\ have investigated the tendency
to phase separation in the Bose-Hubbard model as a more sensitive indicator for
a possible first order transition.  Their preliminary results appear to confirm
the existence of a first order transition in the XSS-SF case, but not for the
CSS-SF transition.  This is in agreement with our RG results, which predict
that the CSS-SF transition remains continuous (3d XY-like) even away from the
multi-critical point.

An interesting open problem is to address the critical behavior in the XSS-SF
case directly in two dimensions more closely, to ascertain whether a
fluctuation-driven first order transition or a continuous transition obtains in
that case for small values of $V$.  This might either be achieved by a
two--loop renormalization group calculation accompanied by some resummation
technique or by a self--consistent calculation, which appears to capture much
of the physics in the mathematically similar problem of the
normal-superconducting transition in zero field\cite{NSzf}.  Whether an
analogous dual formulation of the problem can yield some insight is presently
unclear.

\acknowledgements{It is a pleasure to acknowledge helpful discussions with Anne
  van Otterlo.  We would also like to thank Jan Wilhelm for helping us with
  generating some of the figures. The work of E.F. has been supported by the Deutsche
  Forschungsgemeinschaft (DFG) under contract no. Fr 850/2 and no. SFB 266.
  L.B.'s work was supported at the Institute for Theoretical Physics by grant
  no.  PHY94--07194. We gratefully acknowledge financial support through EU
  Contract ERBCHRX-CT920020 during a workshop at the Institute for Scientific
  Interchange (ISI) in Torino, where part of this work has been done.}

\newpage


\appendix

\section{Mean-Field Decoupling}
\label{decoupling_app}

The long--wavelength action Eq.~(\ref{critical_action}) can be derived
using the method of auxiliary fields.  Because the supersolid phase
exhibits two broken symmetries, it is necessary to introduce two such
fields, which will become order parameters for ODLRO and
crystallinity.  For concreteness, and because the most likely
experimental realization of the SS is in Josephson junction arrays, we
work here with the quantum phase model; the Bose--Hubbard Hamiltonian
may be treated by similar methods.  For simplicity, we also focus here
on the case with only nearest--neighbor interactions, in which case
$\bbox{\psi}$ ordering does not arise.  Standard techniques can be used to
translate the Hamiltonian Eq.~(\ref{phase_hamiltonian}) into the
Euclidean action
\begin{eqnarray}
S_{\rm E} & = & \int \! d\tau \bigg\{ \sum_i \left(i
    n_i\partial_\tau\phi_i -\mu n_i\right) \nonumber \\
    & & - J\sum_{\langle ij\rangle}
    \cos(\phi_i-\phi_j) + \sum_{ij} \frac{1}{2} U_{ij} n_i n_j \bigg\},
\label{pa1}
\end{eqnarray}
where $U_{ij} = U_0\delta_{ij} + U_1\delta_{|i-j|,1}$.  The action enters
the functional integral formulation of quantum mechanics via the
generating functional (partition function)
\begin{equation}
    Z = \int [dn][d\phi] e^{-S_{\rm E}}.
\end{equation}
The Josephson interaction may be decoupled using the
(Hubbard--Stratonovich) identity 
\begin{eqnarray}
    \exp\bigg[ &&\int \! d\tau  \sum_{\langle
    ij\rangle}\cos(\phi_i-\phi_j)\bigg]  \nonumber \\
    &&  = \int [d\bar{\psi}][d\psi]
    \exp\bigg[ -\int\!d\tau \{ \psi_i e^{-i\phi_i} +
    \bar{\psi}_i e^{i\phi_i} \nonumber \\
    & & \;\;\; + (T^{-1})_{ij}\bar{\psi}_i\psi_j\} \bigg],
\end{eqnarray}
where the index sums on the right hand side are implied (we will
adopt this convention in the remainder of this appendix), and the
matrix $T_{ij} \equiv J\delta_{|i-j|,1}$.  As is usually the case with
such decouplings, the functional integral over the $\psi$'s is best
defined by a product measure in momentum space.  To obtain
convergence, the contour for $\tilde{T}(\bbox{q}) < 0$ is then taken
along the imaginary axis.  Before decoupling the
charge interaction, we first separate out the zero momentum component,
according to 
\begin{equation}
U_{ij} = W_{ij} + \bar{U}\delta_{ij},
\label{int_decomp}
\end{equation}
where $\bar{U} = U_0 + 2 d U_1$.  In momentum space, the remaining
coupling is strictly negative,
\begin{equation}
    \tilde{W}({\bf q} \neq 0) = -2dU_0 + 2U_0\sum_{a=1}^d \cos q_a <
    0.
\label{Vtilde}
\end{equation}
We may therefore perform a second Hubbard-Stratonovich to replace
\begin{equation}
    \frac{1}{2} W_{ij} n_i n_j \rightarrow \Phi_i\epsilon_i n_i - 
    \frac{1}{2} (W^{-1})_{ij} \epsilon_i \Phi_i \epsilon_j \Phi_j.
\label{charge_HS}
\end{equation}
We choose to define the auxiliary field with a factor of $\epsilon_i =
(-1)^{\sum_a x_a}$ to bring out the checkerboard correlations.  Note
that because $\tilde{W} < 0$, the functional integral over $\Phi$ is
well behaved (the theory is properly defined by omitting the zero
mode, since $\tilde{W}(\bbox{0}) = 0$).  The last sum can be
rewritten
\begin{equation}
-\frac{1}{2}(W^{-1})_{ij} \epsilon_i \Phi_i \epsilon_j \Phi_j =
\int_{\bbox{q}} 
- \frac{1}{2}|\tilde{\Phi}(\bbox{q})|^2/\tilde{W}(\bbox{q}+\bbox{\pi}),
\end{equation}
where $\tilde{W}(\bbox{q}+\bbox{\pi}) = -2dU_1 - 2U_1\sum_{a=1}^d
\cos q_a \approx -4 d U_1[1- q^2/(4d)]$ for small $q$.  Combining
results, we obtain the full effective action
\begin{eqnarray}
    S_{\rm eff} & = & \int \! d\tau \frac{{d^d\!\bbox{q}}}{{(2\pi)^d}} \bigg\{
    \frac{1}{{8dU_1}}\left( 1 + \frac{{q^2}}{{4d}}\right)|\tilde{\Phi}|^2
    \nonumber \\
    & & + \frac{1}{{2Jd}}\left(1+\frac{{q^2}}{{2d}}\right)|\tilde{\psi}|^2
    \bigg\} + \sum_i s_{1i},
\label{dec_action}
\end{eqnarray}
where we have expanded $\tilde{T}$ around $\bbox{q}=\bbox{0}$, and
introduced the single-particle action
\begin{eqnarray}
    s_{1i} & = & \int \!d\tau \bigg\{ i n_i \partial_\tau \phi_i -
    \alpha_i\cos (\phi_i-\theta_i) \nonumber \\
    & & + \frac{1}{2}\bar{U} n_i^2 - \mu n_i + \Phi_i\epsilon_i n_i
    \bigg\},
\label{sp_action}
\end{eqnarray}
with $\psi_i = -\alpha_i e^{i\theta_i}/2$.  It is useful to first
perform the shift $\phi_i \rightarrow \phi_i + \theta_i$, which
removes the $\theta$ dependence of the cosine at the cost of a
(time-dependent) shift of the chemical potential, $\mu \rightarrow
\tilde{\mu}_i = \mu - i\partial_\tau\theta_i$.  Integration over the
$\phi_i$ and $n_i$ variables can then formally be performed
independently.  It is exactly equivalent to the solution of decoupled
quantum-rotator models, each of which has the single-particle
Hamiltonian
\begin{equation}
    H_1 = \frac{\bar{U}}{2} n^2 - \tilde{\mu} n +
    \Phi\epsilon n - \alpha\cos\phi.
\label{h1def}
\end{equation}
Eq.~(\ref{h1def})\ should be supplemented by the commutation relation
$[\phi,n] = i$.  We will assume slow variations of the phase, so that
$\partial_\tau\theta_i$ is approximately constant.  

The mean--field theory of Refs.~\onlinecite{RS93,vOW94}\ is
recovered in the saddle point approximation, $\delta S_{\rm
eff}/\delta\bar{\psi} = \delta S_{\rm eff}/\delta\Phi = 0$.  This
gives the conditions
\begin{eqnarray}
\alpha_i & = & 4dJ\langle e^{i\phi_i}\rangle, \label{MF1} \\
\Phi_i & = & -4dU_1\epsilon_i \langle n_i\rangle. \label{MF2}
\end{eqnarray}

Because we wish to obtain a Landau theory, allowing for the
possibility of fluctuations, we will explicitly construct the
effective action $S_{\rm eff}$ as a function of $\Phi$ and the
Goldstone mode $\theta$.  It is sufficient to work
perturbatively in $\Phi$.  This is accomplished via the cumulant
expansion
\begin{equation}
    s_{1,\rm eff} = \sum_k \frac{{(-1)^{k+1}}}{{k!}}
    \int_{\tau_1\cdots\tau_k} \langle n(\tau_1)\cdots n(\tau_k) \rangle_C
    \Phi_1 \cdots\Phi_k \epsilon^k,
\label{cumulant}
\end{equation}
where the (cumulant) expectation value is evaluated using the
single--site theory with $\Phi=0$.  For constant $\Phi$, the sum in
Eq.~(\ref{cumulant})\ is just the imaginary time integral of the ground
state energy,  $\int \! d\tau E_1(\Phi)$.

Unfortunately, even these expectation values cannot be evaluated
exactly, due to the form of Eq.~(\ref{h1def}).  As discussed in
Ref.~\onlinecite{RS93}, the one-site Hamiltonian may be rewritten in
the phase representation as that of a particle in a cosine potential.
The associated Schr\"odinger equation is therefore a Mathieu equation,
whose solutions are not available in closed form.  We therefore
concentrate on the region $n_0 \equiv \mu/\bar{U} = 1/2 + \delta$,
with $\delta \ll 1$, in which the checkerboard phase emerges most
strongly, and analytic progress is possible.

The one site problem may be recast in dimensionless form, using
\begin{equation}
\tilde{H}_1 \equiv H_1/\bar{U} = n^2/2 - (1/2 + \lambda)n -
\beta(a+a^\dagger), 
\label{h1dimless}
\end{equation}
where $\lambda = \delta - (i\partial_\tau\theta +
\epsilon\Phi)/\bar{U}$, $\beta = \alpha/(2\bar{U})$, and $a
= e^{i\phi}$.  The operators $a$ and $a^\dagger$ obey the usual
creation and annihilation operator commutation relations with respect
to the number operator, $[a,n] = a$, but {\sl commute} among
themselves $[a,a^\dagger]=0$.  This allows the existence of states
with arbitrarily (positive or negative) integer number, and one should
note that $n \neq a^\dagger a$.  

For small $\delta$, and $\beta=0$,  there are two nearly degenerate
low energy states: $|0\rangle$, with $n=0$ and energy $\tilde{E}_1(n=0) =0$,
and $|1\rangle$, with $n=1$ and energy $\tilde{E}_1(n=1) = -\lambda$ (since
we are eventually going to expand in $\partial_\tau\theta$ and $\Phi$,
$\lambda$ is essentially equal to $\delta$.  All other states (with $n
\neq 0,1$) have the much larger energies $\tilde{E}_1(n \neq 0,1)$ of
order one.  For $\beta \ll 1$, we may therefore work within the
restricted basis of $n=0,1$, in which the Hamiltonian is represented
\begin{equation}
\tilde{H}_1 = -\lambda/2 + \lambda/2\sigma^z - \beta\sigma^x,
\label{h1reduced}
\end{equation}
where $\bbox{\sigma}$ is the vector of Pauli matrices in the standard
representation.  The energy eigenvalues for $\beta \neq 0$ in this
approximation are easily obtained,
\begin{equation}
\tilde{E}_{1\pm} = (-\lambda \pm \Delta)/2,
\label{two_energies}
\end{equation}
where the gap $\Delta = \sqrt{\lambda^2 + 4\beta^2}$.  The two
corresponding (normalized) eigenstates are
\begin{eqnarray}
|-\rangle & = & 
\frac{1}{\sqrt{2\Delta}}\left(\sqrt{\Delta-\lambda}|0\rangle + \frac{{2\beta}}
{\sqrt{\Delta-\lambda}}|1\rangle\right), \\
|+\rangle & = & \frac{1}
{\sqrt{2\Delta}}\left(-\sqrt{\Delta+\lambda}|0\rangle +\frac{{2\beta}}
{\sqrt{\Delta+\lambda}}|1\rangle\right). 
\label{two_states}
\end{eqnarray}
The value of $\beta$ is fixed by the mean--field condition, Eq.~(\ref{MF1}),
equivalent to minimizing the effective Hamiltonian density
\begin{equation}
{\cal H}_{\beta} = \frac{\bar{U}}{2}\left[ - \delta - \Delta_0 +
\frac{\bar{U}}{{Jd}}\beta^2\right],
\end{equation}
where $\Delta_0 = \sqrt{\delta^2 + 4\beta^2}$, and we have taken
$\lambda = \delta$.  Corrections to $\lambda$ will be included via the
Landau expansion.  The minimum occurs when $\beta =
(4d/\bar{U})\sqrt{J^2-J_c^2}$, for $J >J_c = \delta \bar{U}/(8d)$, and
$\beta=0$ for $J <J_c$.  For $\delta=0$, $J_c$ vanishes, because an
arbitrarily weak hopping matrix element breaks the degeneracy between
empty and singly occupied sites.

We are now ready to derive the Landau expansion of the effective
action.  The Hubbard--Stratonovich measure Eq.~(\ref{dec_action})
gives
\begin{eqnarray}
    S_{\rm eff,1} & = & \int \!d^d\bbox{x}d\tau \bigg\{ \frac{1}{{8dU_1}}\Phi^2
    + \frac{1}{{32d^2U_1}}|\nabla\Phi|^2 \nonumber \\
    & &  + \frac{{\bar{U}^2\beta^2}}{{4d^2J}}|\nabla\theta|^2\bigg\},
\label{seff1}
\end{eqnarray}
where we neglected amplitude fluctuations of $\psi$.  A second
contribution is obtained simply by expanding the one-site ground-state
energy in $\Phi$ and $\partial_\tau\theta$,
\begin{eqnarray}
    \Delta S_{\rm eff,2} & = & \int \! d\tau \sum_i - \frac{\bar{U}}{2}
    \Delta_i \nonumber \\
    & & =
    \int \! d^d\bbox{x} d\tau \bigg\{ -\frac{{\beta^2}}
    {{\bar{U}\Delta_0^3}}\Phi^2 + \frac{{\beta^2(\beta^2-\delta^2)}}
    {{\bar{U}^3\Delta_0^7}} \Phi^4 \nonumber \\
    & & +
    \frac{1}{{\bar{U}\Delta_0^3}}(\partial_\tau\theta)^2 -
    \frac{{3\beta^2\delta}}{{\bar{U}^2\Delta_0^5}}
    i\partial_\tau\theta \Phi^2 \bigg\},
\label{raw_eff_action}
\end{eqnarray}
up to constants.  The only remaining non-trivial term is
$(\partial_\tau\Phi)^2$.  It cannot be calculated from the ground
state energy $E_1$, because it arises from the time dependence in
$H_1$, for which stationary quantum mechanics is not valid.  It can,
however, be calculated from the more general cumulant expansion,
Eq.~(\ref{cumulant}), which also provides a check on the $\Phi^2$
coefficient in Eq.~(\ref{raw_eff_action}).  Taking $k=2$ in
Eq.~(\ref{cumulant})\ gives
\begin{eqnarray}
\Delta S_{\rm eff} & = & -\frac{1}{2} \sum_i \int_{\tau,\tau'} \langle
n(\tau)n(\tau') \rangle_C \Phi_i(\tau)\Phi_i(\tau') \nonumber \\
& & = - \int \! \frac{{d\omega}}{{2\pi}} \frac{{C(\omega)}}{2}
|\Phi(\omega)|^2, 
\label{cumulant_result}
\end{eqnarray}
where we used time translation invariance to express the result in
terms of $C(\omega) \equiv \int \! d\tau \langle n(\tau)n(0)\rangle_C
e^{i\omega\tau}$.  The imaginary time correlation function is easily
expressed quantum mechanically as
\begin{eqnarray}
C(\tau) & = & \langle -| n e^{-(H-E_{1-})|\tau|}n |-\rangle - \langle
-|n|-\rangle^2 \nonumber \\
& & = 
\frac{\beta^2}{\Delta_0^2}e^{-\bar{U}\Delta|\tau|},
\label{Ctau}
\end{eqnarray}
after inserting the complete set (in this approximation)
$|+\rangle\langle+| + |-\rangle\langle-|$, and using
Eqs.~\ref{two_energies}--\ref{two_states}.  Fourier transforming gives
$C(\omega) = (2\beta^2/\Delta_0)[\bar{U}/(\bar{U}^2\Delta_0^2 +
\omega^2)]$.  For $\omega=0$, Eq.~(\ref{cumulant_result})\ then agrees
with the $\Phi^2$ term calculated earlier, and provides the
``kinetic'' correction
\begin{equation}
\Delta S_{\rm eff,3} = \int \! d^d\bbox{x} d\tau  
\frac{\beta^2}{{\bar{U}^3\Delta_0^5}} |\partial_\tau\Phi|^2.
\label{kinetic_term}
\end{equation}

The full effective action is thus $S_{\rm eff} = S_{\rm eff,1} + S_{\rm eff,2}
+ S_{\rm eff,3}$.  It is brought into the standard form of
Eqs.~\ref{critical_action}--\ref{ptcoupling} by the rescaling
\begin{equation}
    \Phi \rightarrow 4d\sqrt{U_1}\Phi.
\end{equation}
After rescaling, the Landau parameters are
\begin{eqnarray}
t & = & 4d - \frac{{32d^2\beta^2 U_1}}{{\Delta_0^3\bar{U}}}, \\
u & = & 6144 d^4 \frac{{\beta^2(\beta^2-\delta^2)}}{{\Delta_0^7}}
\frac{U_1^2}{\bar{U}^3}, \\
c^2 & = & \frac{{\Delta_0^5}}{{32d^2\beta^2}}\frac{\bar{U}^3}{U_1}, \\
v^2 & = & \frac{{\beta^2\Delta_0^3}}{{4d^2}}\frac{\bar{U}^3}{J}, \\
\sigma & = & - \frac{{48d^2\beta^2\delta}}{{\Delta_0^5}}
\frac{U_1}{\bar{U}^2}, \\
\frac{\rho_s}{m^2} & = & \frac{\beta^2}{{2d^2}}\frac{\bar{U}^2}{J}.
\end{eqnarray}
{From} these equations, we can determine the bare velocity ratio $V = v/c$, as
\begin{equation}
V^2 = 32 d^2 \left( \frac{J}{\bar{U}} \right) \frac{U_1}{\bar{U}},
\end{equation}
at the multi-critical point $\delta = 0$.  These results are valid for $\beta
\sim J/\bar{U} \ll 1$, so that $V \ll 1$ (since $U_1 < \bar{U}$).  In this
limit the $V \ll 1$ approximation used in section \ref{SF_X_transition}\ is
valid even in the early stages of renormalization (i.e. uniformly at all
scales).

\section{Lack of Momentum-Dependent Loop Contribution to $\Phi$ Propagator}
\label{loop_appendix}

In this appendix we consider the integral
\begin{equation}
I(k,\omega) \equiv \int_{\bbox{p},\nu} \frac{\nu^2}{{p^2 + \nu^2/v^2}}
\frac {e^{-|p|a-|\bbox{p}+\bbox{k}|a}}
      {{(\bbox{p}+\bbox{k})^2 + (\nu+\omega)^2}},
\label{Idef}
\end{equation}
where $\int_{\bbox{p},\nu} \equiv \int \! \frac{{d^3\bbox{p}}}{{(2\pi)^3}}
\frac{{d\nu}}{{2\pi}}$, and $a=1/\Lambda$.  We have taken $d=3$ to obtain the
leading term of $O(\varepsilon^0)$, chosen an exponential cut-off for momenta,
and set $c=1$.  We will also make use of the limit $v \rightarrow 0$, to which
the theory is driven (within the $\varepsilon$-expansion) by the monotonic
negative velocity renormalization.  It is crucial here that {\sl both} internal
lines are cut-off: physically, we define the theory unambiguously by confining
{\sl all} momenta of the theory to be less than $\Lambda$.  In the momentum
shell RG, we would perform an analogous integral iteratively in order to reduce
$\Lambda$.  Here we will instead perform the entire integral with fixed
cut-off.  It may then be differentiated with respect to $\Lambda$ to yield the
appropriate contribution from a particular scale.  Any renormalizations must
thus appear as logarithmic divergences, i.e. $\ln(\Lambda/k)$ factors in the
integral.  In particular, the coefficients of $k^2\ln(\Lambda/k)$ and
$\omega^2\ln(\Lambda/k)$ determine the renormalization of the $|\nabla\Phi|^2$
and $(\partial_\tau\Phi)^2$ terms in the action, respectively.

Focusing first on the gradient term, consider $I_0(k) \equiv
I(k,\omega=0)$.  The frequency integral may be simply performed by
contour integration to yield
\begin{equation}
I_0(k) = \frac{v^2}{2} \int_{\bbox{p}_1,\bbox{p}_2} \! \frac{1}{{p_2 +
vp_1}} e^{-p_1 a - p_2 a} (2\pi)^3\delta(\bbox{p}_1+\bbox{p}_2+\bbox{k}),
\label{I1}
\end{equation}
where we have inserted a delta function to put the momenta of the two
propagators on equal footing.  Using the Fourier representation
$(2\pi)^3\delta(\bbox{k})= \int_{\bbox{x}} \exp[i\bbox{k}\cdot\bbox{x}]$ and
going to angular coordinates $y=\cos(\theta)$, where $\theta$ is the azimuthal
angle, we have
\begin{eqnarray}
I_0 & = & \frac{v^2}{2}\int_{\bbox{x}}
\frac{{e^{i\bbox{k}\cdot\bbox{x}}}}{{(2\pi)^4}} 
\int^\Lambda\!dp_1 dp_2 (p_1p_2)^2 \int_{-1}^1\!
dy_1 dy_2 \nonumber \\
& & \frac{1}{{p_2 + vp_1}}e^{ix(p_1y_1+p_2y_2)},
\label{I2}
\end{eqnarray}
where $\int^\Lambda dp \equiv \int_0^\Lambda dp e^{-pa}$.  Performing the
$y$--integrals gives
\begin{equation}
I_0 = \frac{v^2}{{2\pi^3k}}\int_0^\infty \! dx \frac{{\sin kx}}{x}
\int^\Lambda\! dp_1 dp_2 p_1 p_2 \frac{{\sin p_1 x \sin p_2 x}}{{p_2 +
vp_1}}.
\label{I3}
\end{equation}
The $v \rightarrow 0$ limits of the $p$--integrals are well defined.  Taking it
gives
\begin{equation}
I_0 = \frac{v^2}{{\pi^3k}} \int_0^\infty \! dx 
\frac{{ax\sin kx}}{{(a^2 + x^2)^3}}.
\label{I4}
\end{equation}
The $k^2$ term is
\begin{equation}
    \frac{{dI_0}}{{dk^2}} \; \matrix{\longrightarrow \cr {k \rightarrow
    0}} \; - \frac{v^2}{{32\pi^2}}.
\label{I5}
\end{equation}
Note that this is a {\sl finite} result, with no $\ln(ka)$ dependence.  Thus
there is no renormalization (in the scaling sense) of the
$|\bbox{\nabla}\Phi|^2$ term.

We now consider possible renormalization of the $|\partial_\tau\Phi|^2$ term.
It is sufficient in this case to compute $I' = \partial I/\partial
\omega^2|_{\omega=0}$ for small $k$.  A divergent (scaling) renormalization
should appear in $I'$ again as a $\ln(ka)$, so $k$ must be kept non-zero until
the end of the calculation.  Differentiating Eq.~(\ref{Idef})\ gives
\begin{eqnarray}
    I' & = & v^2\int_{\bbox{p}_1,\bbox{p}_2,\nu} \! \left[1 - 
           \frac{{v^2 p_1^2}}{{\nu^2+v^2p_1^2}}\right] 
           \bigg[\frac{3}{{(\nu^2+p_2^2)^2}} \nonumber \\
       & &   - \frac{{4p_2^2}}{{(\nu^2 + p_2^2)^3}} \bigg]  
             e^{-p_1 a - p_2 a} (2\pi)^3\delta(\bbox{p}_1+\bbox{p}_2+\bbox{k}).
\label{Ip1}
\end{eqnarray}
It is convenient to take the $v \rightarrow 0$ limit in the integrand at this
stage, which allows one to drop the second term in the first set of square
brackets in Eq.~(\ref{Ip1}).  Performing the frequency integrals then gives
\begin{eqnarray}
I' & = & v^2 \int_{\bbox{p}_1,\bbox{p}_2} \! \left[ \frac{3}{{4p_2^3}} -
4p_2^2 \frac{3}{{16 p_2^5}}\right] \nonumber \\
& \times & e^{-p_1 a - p_2 a}
(2\pi)^3\delta(\bbox{p}_1+\bbox{p}_2+\bbox{k}),
\label{Ip2}
\end{eqnarray}
which is identically zero!  The $|\partial_\tau\Phi|^2$ term is therefore
unrenormalized.


\section{Perturbation Theory for the Vertex Functions}
\label{app_pert}

This appendix comprises the Feynman diagrams to one--loop order for the
effective free energy functional of the supersolid to superfluid transition.
We start with a list of the contributions to one--loop order to the fully
wavevector-- and frequency--dependent two--point vertex functions.  The Feynman
diagrams for the vertex functions $\Gamma_{ln}$, with $l$ fields of the
superfluid phase and $n$ Ising fields, are depicted in
Figs.~(\ref{gamma_02})-(\ref{gamma_04}).  The corresponding analytical
expressions for $\Gamma_{02} ({\bf q},\omega)$ is given by
\begin{equation}
   \Gamma_{02}^{(a)} ({\bf q}=0,\omega=0) =
   \frac{u}{2} \int_{{\bf p},\mu} \frac{1}{t + (\mu/c)^2 + p^2} \, ,
\label{a1}
\end{equation}
and
\begin{eqnarray}
   \Gamma_{02}^{(b)} ({\bf q},\omega) = 
      &&4 (\sigma c v)^2  \int_{{\bf p},\mu}
      \frac{1}
           {(\omega + \mu)^2 + c^2[t+({\bf p}+{\bf q})^2]}
      \nonumber \\
      &&\quad\times \frac{\mu^2}{[\mu^2 + v^2 p^2]}
\label{a2}
\end{eqnarray}
Performing the frequency integrals using the residue theorem we find
\begin{equation}
   \Gamma_{02}^{(a)} ({\bf q}=0,\omega=0) =
   - t \, \frac{u c}{8} \, \frac{S_d}{\varepsilon}  \, ,
\label{a3}
\end{equation}
where we have subtracted the $T_c$--shift 
$\Gamma_{02}^{(a)} ({\bf 0}, 0) \mid_0$ and 
\begin{eqnarray}
   \Gamma_{02}^{(b)}  ({\bf q},\omega) =
   && \frac{4 (\sigma v c)^2}{c}  \int_{{\bf p}}
   \Biggl\{ \frac{1}{2A} - \left(\frac{v}{c}\right)^2  
   ({\bf p}+{\bf q})^2 \frac{A+B}{2 AB}
   \nonumber \\
   &&\quad \times \frac{1}{(\omega/c)^2 + (A+B)^2} \Biggr\} \, ,
\label{a4}
\end{eqnarray}
where $A= \sqrt{t+p^2}$ and $B = (v/c) \sqrt{({\bf p}+{\bf q})^2}$.  The
wavevector integral of Eq.~(\ref{a4}) can be evaluated for several limits
listed below:

\noindent (a) $\omega = 0$, ${\bf q} = {\bf 0}$, $t \neq 0$: 
\begin{equation}
  \Gamma_{02}^{(b)} ({\bf 0}, 0) =
     - t (v \sigma)^2 c \frac{S_d}{\varepsilon}
       \frac{1}{(1+v/c)^2} \, ,
\label{a5}
\end{equation}
where we have again subtracted the corresponding $T_c$--shift 
$\Gamma_{02}^{(b)} ({\bf 0}, 0) \mid_0$.

\noindent (b) $\omega = 0$, ${\bf q} \neq {\bf 0}$, $t = 0$: 
\begin{equation}
  \Gamma_{02}^{(b)} ({\bf q}, 0) =
    - q^2 v^3 \sigma^2  \frac{S_d}{\varepsilon} 
   \frac{1}{(1+v/c)^3} {\cal A} (v/c,d) \, ,
\label{a6}
\end{equation}
with 
\begin{equation}
  {\cal A}(v/c,d) = \frac{8-2d+(v/c)(9-3d)+(v/c)^2(3-d)}{d} 
\label{aA}
\end{equation}
\noindent (c) $\omega \neq 0$, ${\bf q} = {\bf 0}$, $r = 0$: 
\begin{eqnarray}
  \Gamma_{02}^{(b)} ({\bf q}, 0) =
    \left(\frac{\omega}{c}\right)^2 
    2 v^3 \sigma^2 \frac{S_d}{\varepsilon}
    \frac{1}{(1+v/c)^3} \, .
\label{a7}
\end{eqnarray}
The analytical expression for $\Gamma_{20} ({\bf q},\omega)$ at $\omega = 0$
reads
\begin{equation}
   \Gamma_{20} ({\bf q},0) =
   2 \omega^2 \sigma^2 c \int_{{\bf p},\mu}
   \frac{1}{[t + \mu^2 + p^2]^2} \, ,
\label{a8}
\end{equation}
and after performing the frequency integrals using the residue theorem we find
\begin{equation}
   \Gamma_{20} ({\bf q},0) =
   \frac{\omega^2 c}{2} \sigma^2 \frac{S_d}{\varepsilon} \, .
\label{a9}
\end{equation}
The vertex function $\Gamma_{12} ( {\bf q},0) := \Gamma_{12} ({\bf
  q},\omega;-{\bf q}/2,0;-{\bf q}/2,0)$ to one--loop order has two
contributions
\begin{equation}
   {\Gamma_{12}^{(a)}} ( {\bf q},0) =
    - \omega \sigma u  
    \int_{{\bf p},\mu} \frac{1}{[t+(\mu/c)^2+p^2]^2} \, ,
\label{a10}
\end{equation}
and
\begin{equation}
   {\Gamma_{12}^{(b)}} ( {\bf q},0) =
   - 8 \omega \sigma^3
   \int_{{\bf p},\mu}
   \frac{\mu^2}{[t+(\mu/c)^2+p^2]^2 [(\mu/v)^2 + p^2]} \, .
\label{a11}
\end{equation}
After performing the frequency integrals we find
\begin{equation}
   {\Gamma_{12}^{(a)}} ( {\bf q},0) =
   - \omega \sigma u c \frac{S_d}{\varepsilon}
\label{a12}
\end{equation}
\begin{equation}
   {\Gamma_{12}^{(b)}} ( {\bf q},0) =
   - 2 c \omega \sigma^3 v^2 \frac{S_d}{\varepsilon}
    \frac{1}{(1+v/c)^2}
\label{a13}
\end{equation}
The vertex function $\Gamma_{04}$ at zero external frequencies and momenta has
three contributions.
\begin{equation}
   \Gamma_{04}^{(a)} =
   - \frac{3}{2} u^2 c 
   \int_{{\bf p},\mu}
   \frac{1}{[t+(\mu/c)^2+p^2]^2} \, , 
\label{a14}
\end{equation}
\begin{equation}
   \Gamma_{04}^{(b)} =
   - 24 u \sigma^2
   \int_{{\bf p},\mu}
   \frac{\mu^2}{[t+(\mu/c)^2+p^2]^2 [(\mu/v)^2 + p^2]} \, , 
\label{a15}
\end{equation}
\begin{equation}
   \Gamma_{04}^{(c)} =
   -96 \sigma^4 
   \int_{{\bf p},\mu}
   \frac{\mu^4}
   {[t+(\mu/c)^2+p^2]^2 [(\mu/v)^2 + p^2]^2} \, .
\label{a16}
\end{equation}
After performing the frequency integrals we find
\begin{eqnarray}
   \Gamma_{04}^{(a)} &=&
   - \frac{3 u^2 c}{8} \frac{S_d}{\varepsilon}
\label{a17} \\
   \Gamma_{04}^{(b)} &=&
   - 6 u c (\sigma v)^2 \frac{S_d}{\varepsilon} \frac{1}{(1+v/c)^2} \, , 
\label{a18} \\
   \Gamma_{04}^{(c)} &=&
   - 24  (\sigma v)^4 c \frac{S_d}{\varepsilon} \frac{1}{(1+v/c)^3}  \, .
\label{a19}
\end{eqnarray}


\newpage

\begin{table}
\setdec 0.00
\caption{Transformation properties of the checkerboard $\Phi$ and collinear
order parameter $\Psi$ under coordinate transformations.}
\vspace{0.5cm}
\begin{tabular}{ccc}
Symmetry & Coord. Transf. & Field Transf. \\
\tableline
Translation & $x_i \rightarrow x_i + 1$ & $\Phi \rightarrow - \Phi$ \\
& & $\psi_i \rightarrow -\psi_i$ \\
Reflection & $x_i \rightarrow - x_i$ & Invariant \\
$\pi/2$ Rotation & $x_i \rightarrow x_j$ & $\psi_i \leftrightarrow
\psi_j$ \\
& $x_j \rightarrow - x_i$ & \\
Time Reversal & $\tau \rightarrow - \tau$ & $\theta \rightarrow
-\theta$ \\
\end{tabular}
\label{symmetrytable}
\end{table}

\newpage

\begin{figure}
\narrowtext
\epsfxsize=\columnwidth\epsfbox{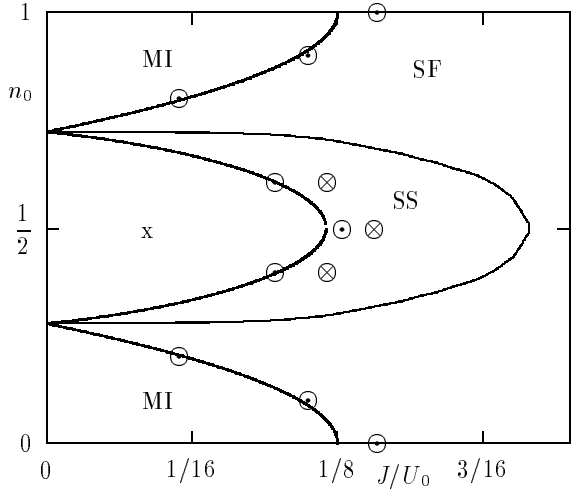}
\vspace{15pt}
\caption{Phase diagrams for soft-core bosons, as obtained from the mean-field
  analysis \protect\cite{RS93,Wag94} of the Quantum Phase Hamiltonian with
  on-site and nearest-neighbor ($U_{1}/U_{0}=1/5$) interaction, $n_0 = \mu /
  \sum_i U_{ij}$. The symbols are the Monte Carlo data as discussed in
  Ref.~\protect\cite{vOW94,vOW95}.  The checkerboard charge-density wave is
  denoted by ``X'', the supersolid phase by ''SS'', the superfluid phase
  by ``SF'' and the Mott-insulating phase by ``MI''. Taken from
  Ref.~\protect\cite{vOW95}.}
\label{phase_diagram}
\end{figure}

\begin{figure}
\epsfxsize=0.9\columnwidth\epsfbox{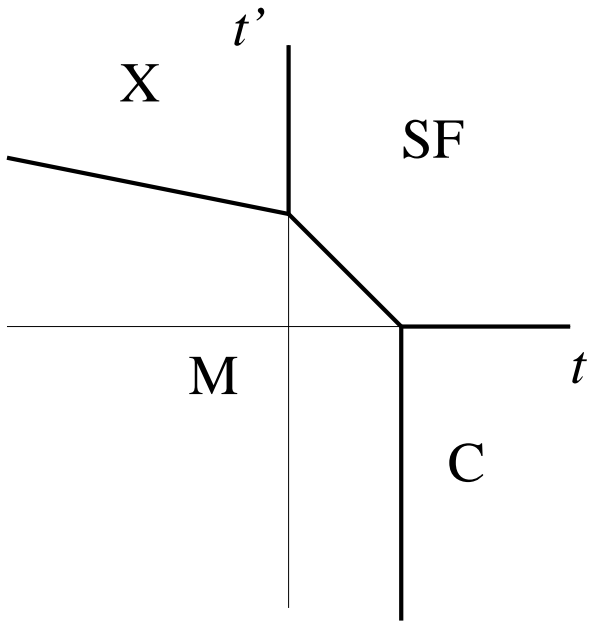}
\vspace{15pt}
\caption{Topology of the mean-field phase diagram.}
\label{MF_phase_diagram_fig}
\end{figure}

\begin{figure}
\epsfxsize=\columnwidth\epsfbox{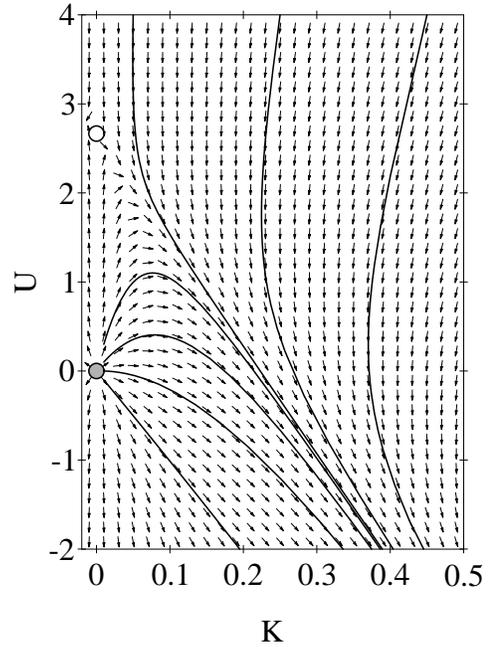}
\vspace{15pt}
\caption{Momentum-shell flow diagram at the SF-X critical point for $V=0$ in
$d=2$ dimensions. The Ising I fixed point and the Gaussian I fixed point are 
indicated by the open and shaded circle, respectively.}
\label{mom_shell_flow_fig}
\end{figure}

\begin{figure}
\epsfxsize=0.7\columnwidth\epsfbox{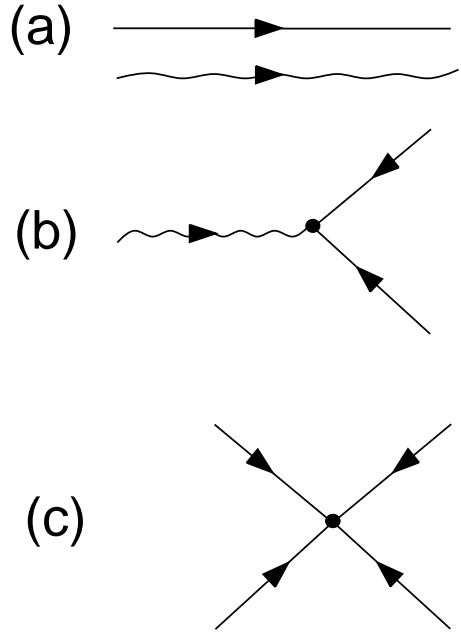}
\vspace{15pt}
\caption{Elements of perturbation theory. (a) Propagators for the Ising field
  $\Phi$ (solid line) and the superfluid phase $\Theta$ (wiggly line). (b)
  Vertex for the coupling between the superfluid phase and the crystalline
  order parameter. (c) $\Phi^4$ term for the Ising field $\Phi$.}
\label{elements_perturbation_theory}
\end{figure}

\begin{figure}
\epsfxsize=0.9\columnwidth\epsfbox{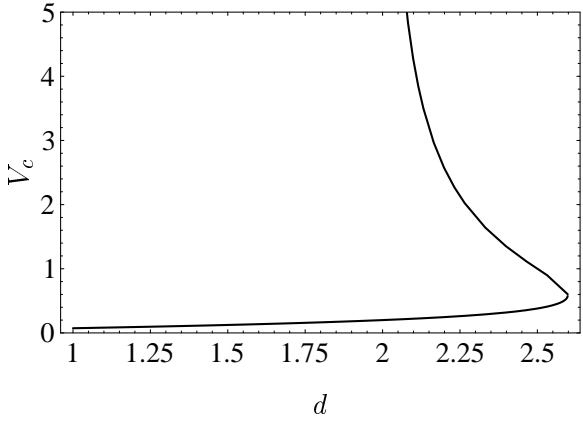}
\vspace{15pt}
\caption{Critical values $V_c^{1,2}$ (lower and upper curve) for the ratio 
  $V=v/c$ as a function of the dimension $d$. For $V<V_c^1$ the flow of $V$
  tends to $V=0$ and one gets runaway trajectories for the flow of the coupling
  constant. For $V_c^2>V>V_c^1$ the flow of $V$ tends to $V_c^2$, and finally
  for $V>V_c^2$ the flow of $V$ tends to $V=\infty$.}
\label{v_crit}
\end{figure}

\begin{figure}
\epsfxsize=\columnwidth\epsfbox{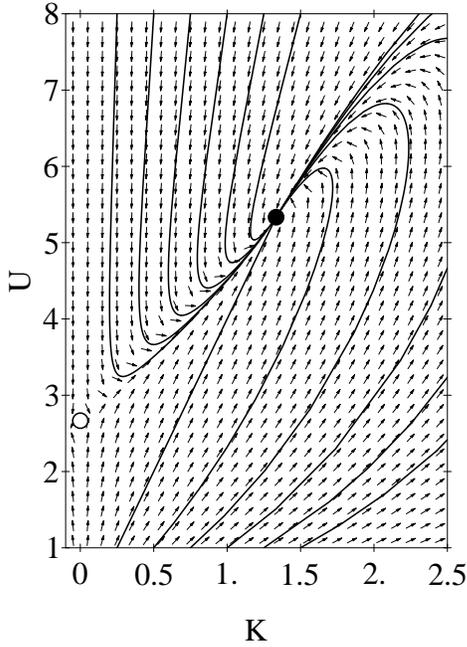}
\vspace{15pt}
\caption{Flow diagram of the SF-X transition in the limit $V = \infty$ for 
  $d=2$. The topology of the flow is determined by the presence of the unstable
  Ising fixed point at $(K,U) = (0,8/3)$ (open circle) and the stable
  non-Bose liquid fixed point at $(K,U) = (4/3,16/3)$ (filled circle).}
\label{flow_non_bose_liquid}
\end{figure}

\begin{figure}
\epsfxsize=0.6\columnwidth\epsfbox{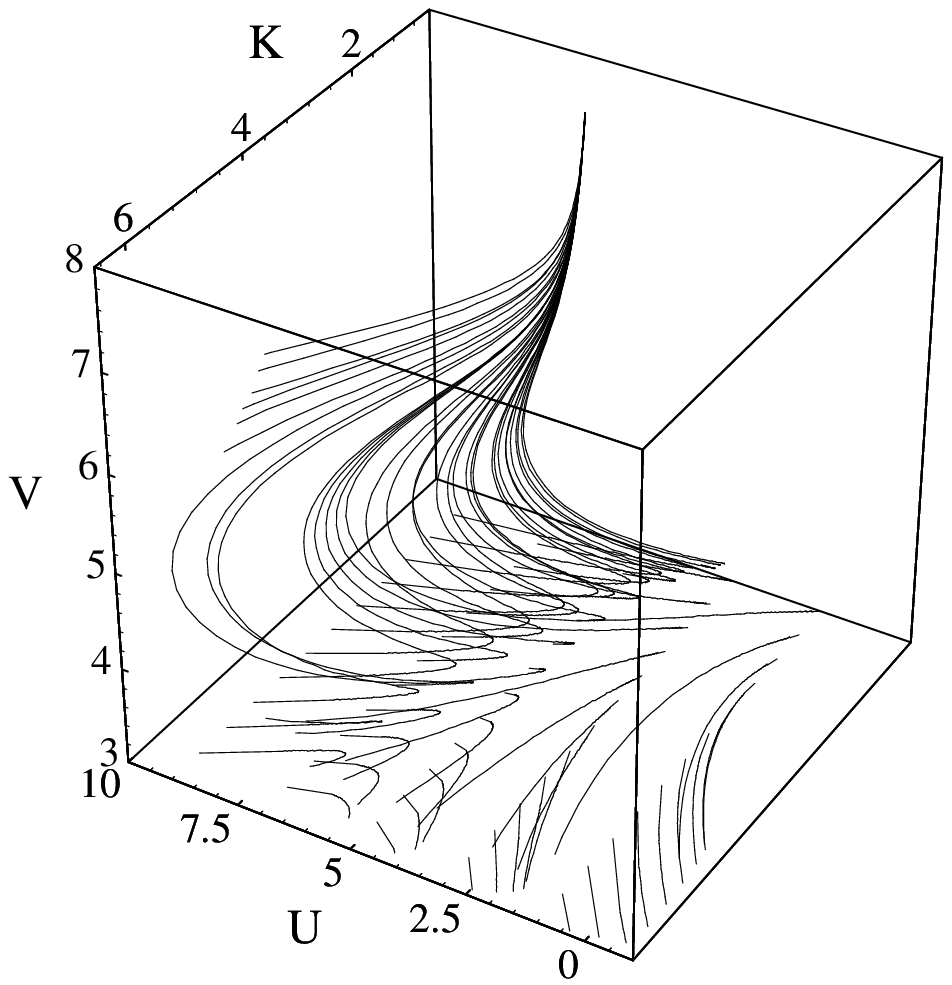}
\epsfxsize=0.6\columnwidth\epsfbox{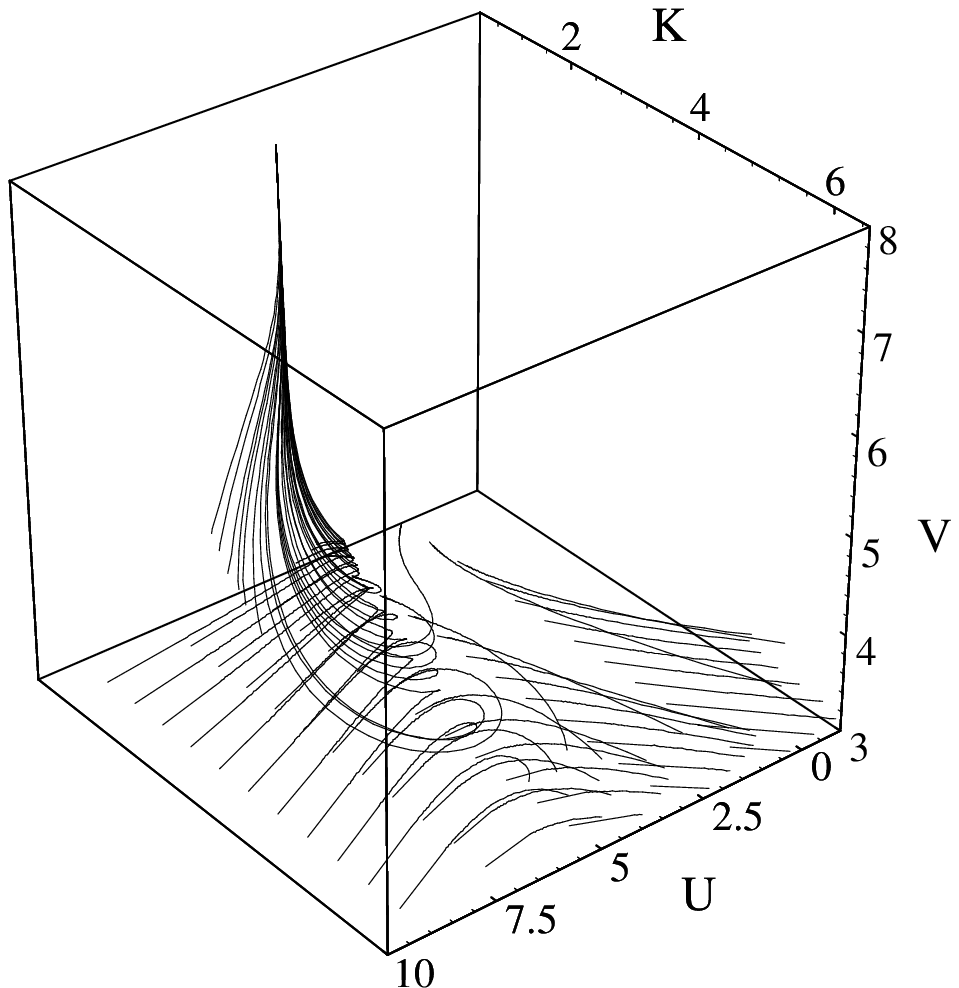}
\vspace{15pt}
\caption{Two different views of the RG--trajectories from the RG--equations
  Eqs.~(\protect\ref{rg1})--(\protect\ref{rg3}). }
\label{figure5}
\end{figure}

\begin{figure}
{\epsfxsize=0.6\columnwidth\epsfbox{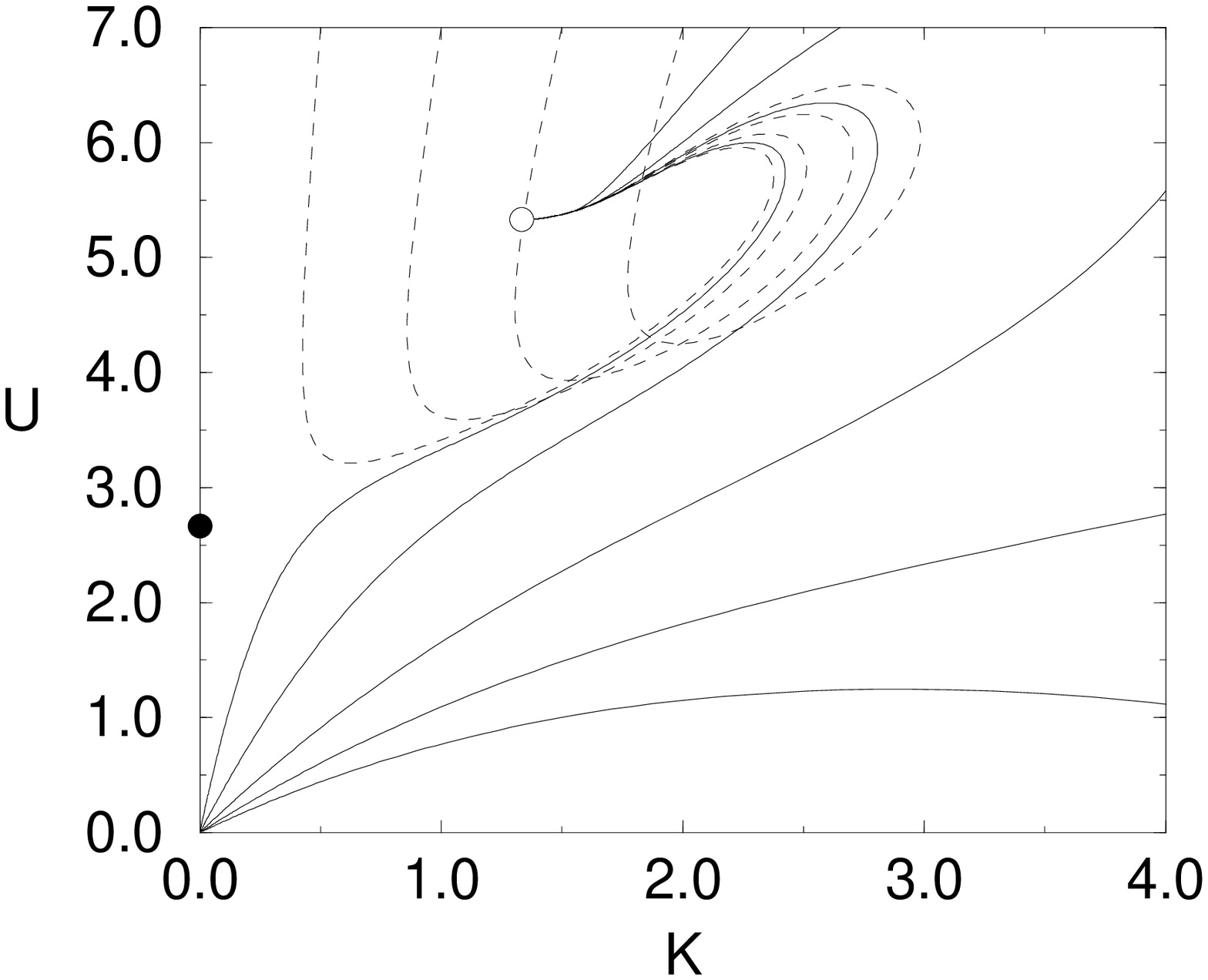}}
\vspace{5pt}
{\epsfxsize=0.6\columnwidth\epsfbox{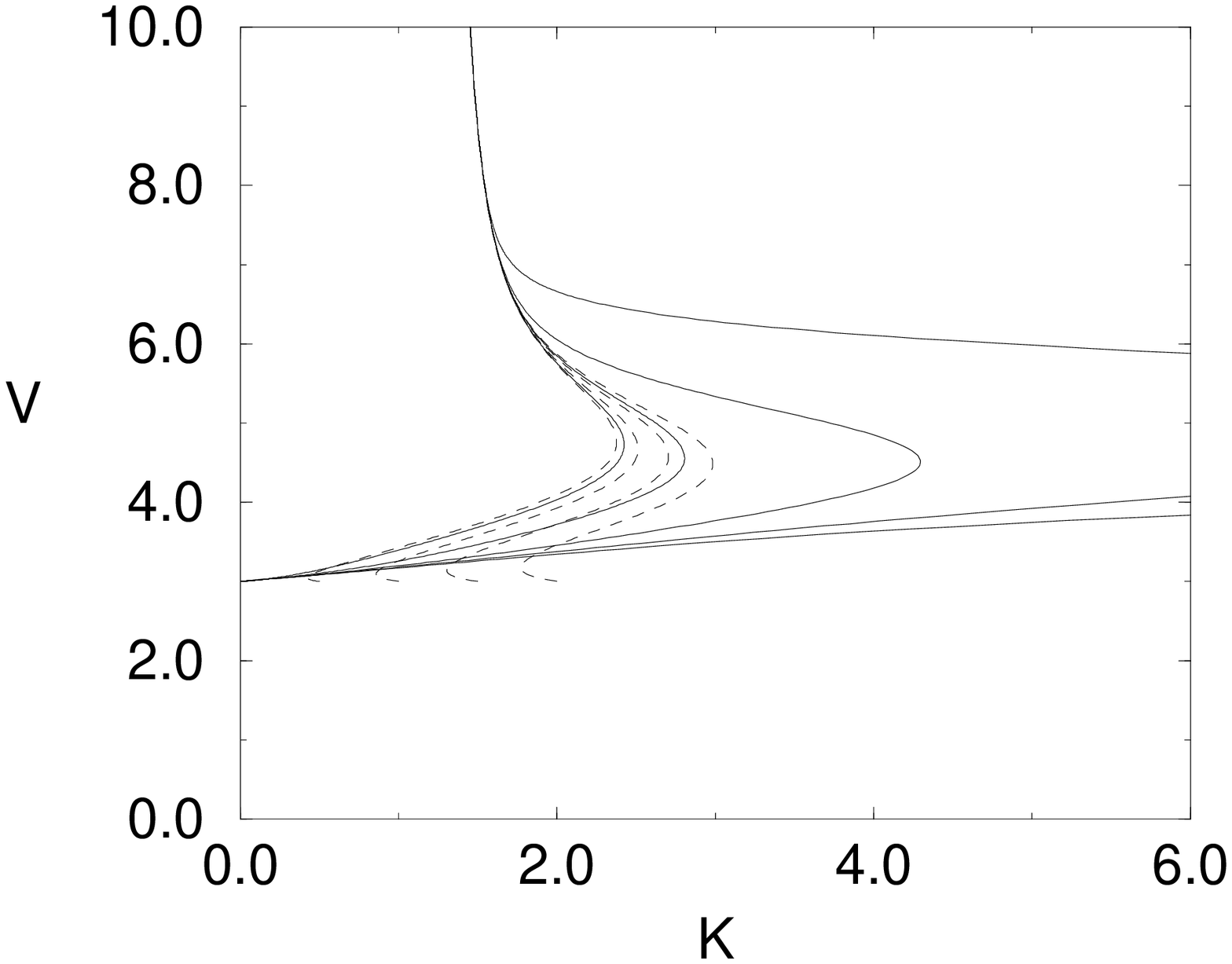}}
\vspace{15pt}
\caption{Two--dimensional projections of the three-\--di\-men\-sional 
  $(U,K,V)$--flow diagram. The trajectories all start at $V=3$ and a set of
  values for $U$ and $K$. The first set (dashed lines) is $U=7$ and
  $K=0.5,1.0,1.5,2.0$, the second set (solid lines) is $U=0.01$ and
  $K=0.001,0.005,0.0075,0.01$.}
\label{figure6}
\end{figure}

\begin{figure}
{\epsfxsize=0.6\columnwidth\epsfbox{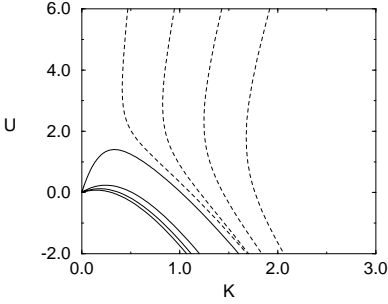}}
\vspace{5pt}
{\epsfxsize=0.6\columnwidth\epsfbox{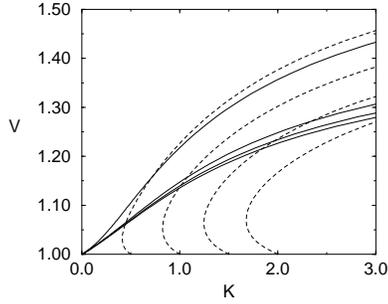}}
\vspace{15pt}
\caption{Two--dimensional projections of the three-\--di\-men\-sional 
  $(U,K,V)$--flow diagram. The trajectories all start at $V=1$ and a set of
  values for $U$ and $K$. The first set (dashed lines) is $U=7$ and
  $K=0.5,1.0,1.5,2.0$, the second set (solid lines) is $U=0.01$ and
  $K=0.001,0.005,0.0075,0.01$.}
\label{figure7}
\end{figure}

\begin{figure}
\epsfxsize=0.6\columnwidth\epsfbox{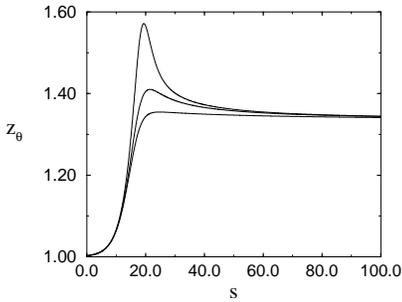}
\vspace{5pt}
\epsfxsize=0.6\columnwidth\epsfbox{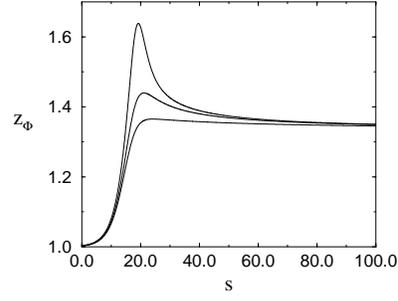}
\vspace{15pt}
\caption{Effective dynamic critical exponents a) $z_{\Theta} (s)$ and b) 
  $z_{\Phi}$ for the initial values $U(s = 1) = 8/3$, $K(s = 1) = 1/100$
  and a series of rations of sound velocities $V(1) = 10^{-k}$ with
  $k=3,2,...,-1$.}
\label{figure8}
\end{figure}

\begin{figure}
\epsfxsize=0.9\columnwidth\epsfbox{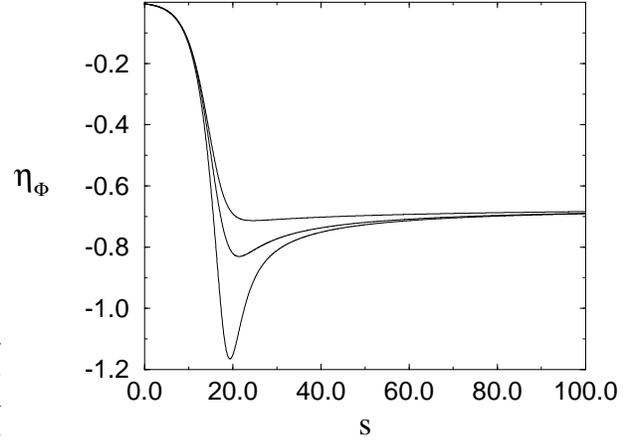}
\vspace{15pt}
\caption{Effective Fisher exponent $\eta_{\Phi} (s)$
  for the initial values $U(s = 1) = 8/3$, $K(s = 1) = 1/100$ and a series of
  rations of sound velocities $V(1) = 10^{-k}$ with $k=3,2,...,-1$.}
\label{figure9}
\end{figure}

\begin{figure}
\epsfxsize=0.9\columnwidth\epsfbox{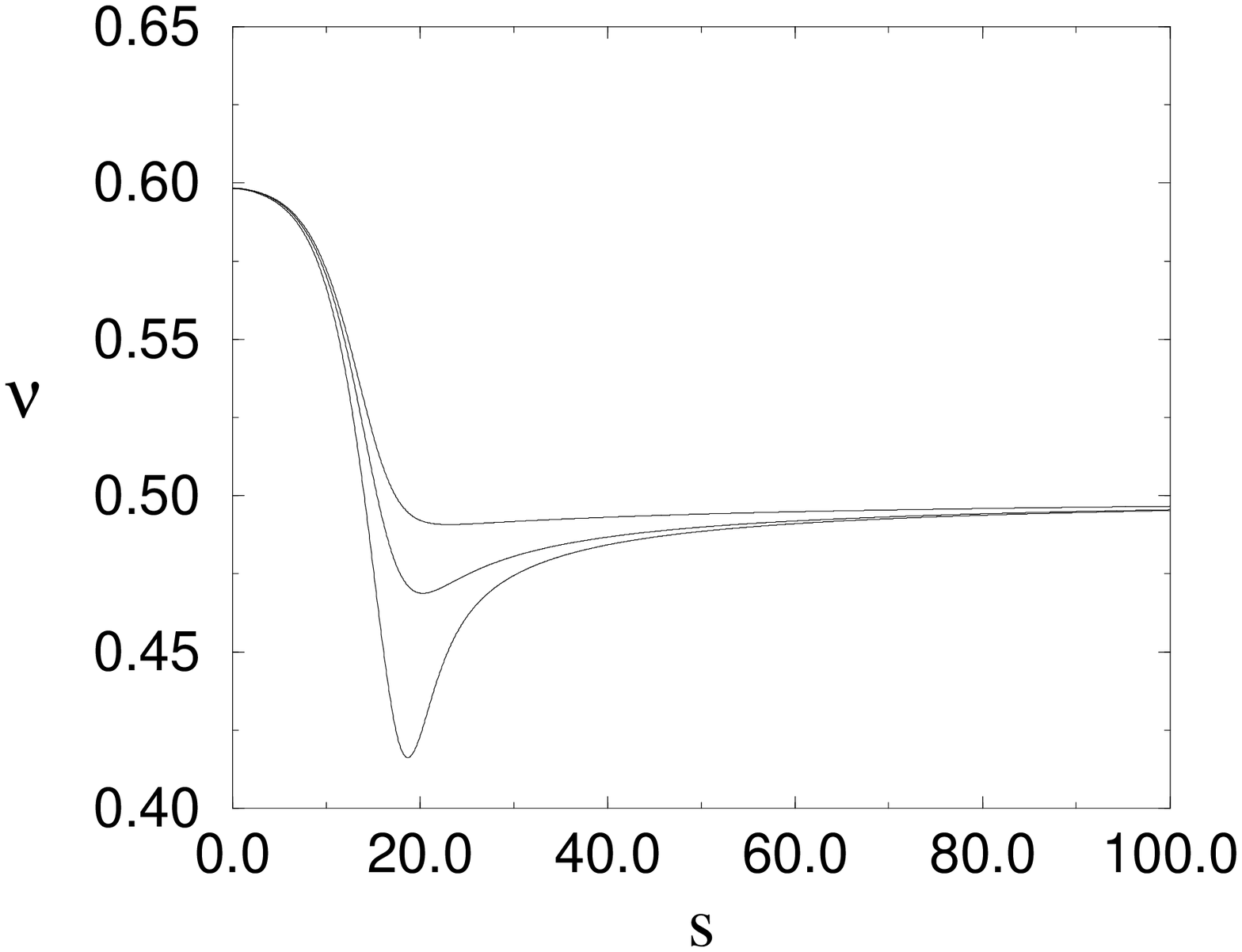}
\vspace{15pt}
\caption{Effective correlation length exponent $\nu (s)$ for the
initial values $U(s = 1) = 8/3$, $K(s = 1) = 1/100$ and a series of
ratios of sound velocities $V(1) = 10^{-k}$ with $k=3,2,...,-1$.}
\label{figure10}
\end{figure}

\newpage

\begin{figure}
\rotate[r]{\epsfysize=0.9\columnwidth\epsfbox{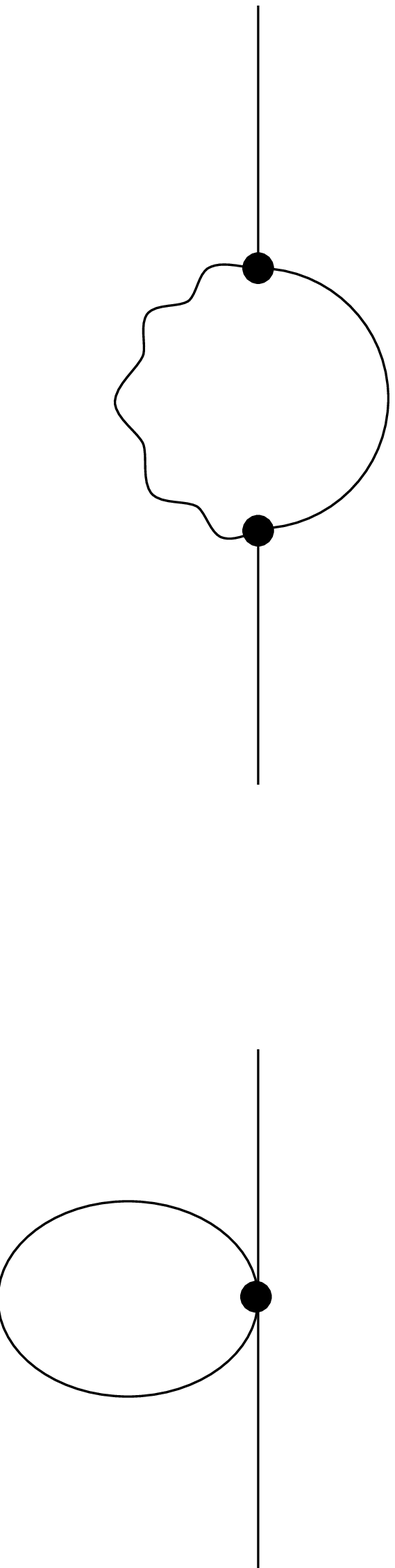}}
\vskip 0.5truecm
\caption{Feynman diagrams for the dynamic perturbation expansion of
  $\Gamma_{02}$ to one--loop order.}
\label{gamma_02}
\end{figure}

\begin{figure}
\rotate[r]{\epsfysize=0.5\columnwidth\epsfbox{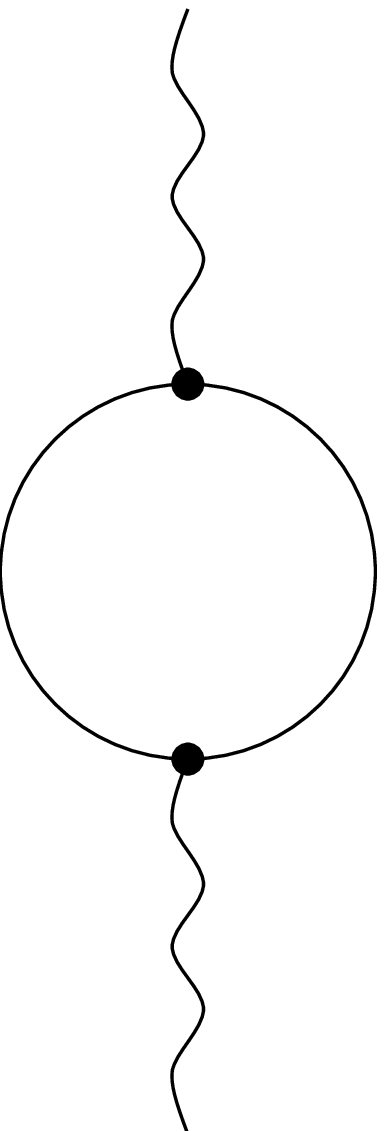}}
\vskip 0.5truecm
\caption{Feynman diagrams for the dynamic perturbation expansion of
  $\Gamma_{20}$ to one--loop order.}
\label{gamma_20}
\end{figure}

\begin{figure}
\rotate[r]{\epsfysize=0.9\columnwidth\epsfbox{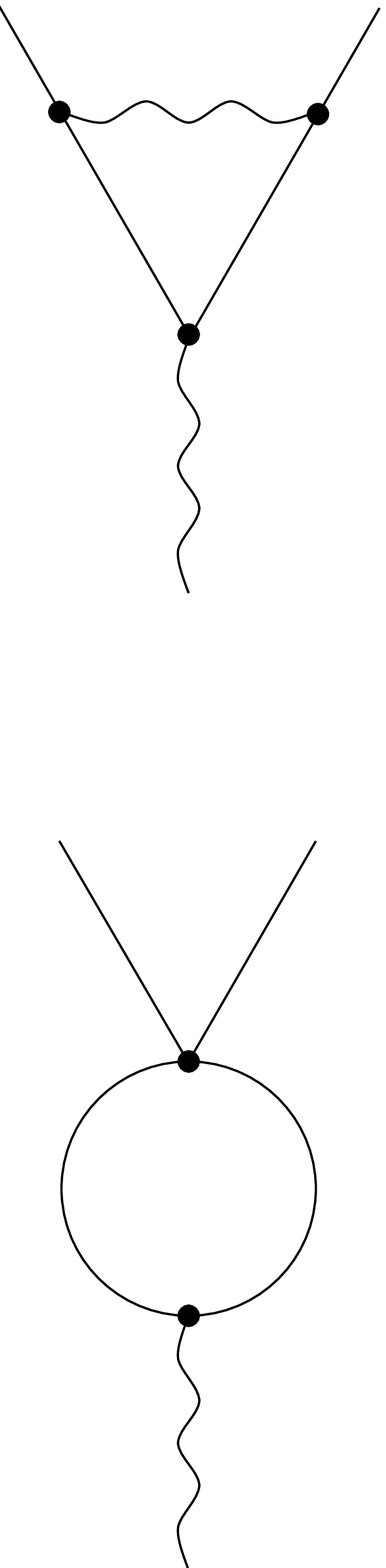}}
\vskip 0.5truecm
\caption{Feynman diagrams for the dynamic perturbation expansion of
  $\Gamma_{12}$ to one--loop order.}
\label{gamma_12}
\end{figure}

\begin{figure}
\rotate[r]{\epsfysize=0.9\columnwidth\epsfbox{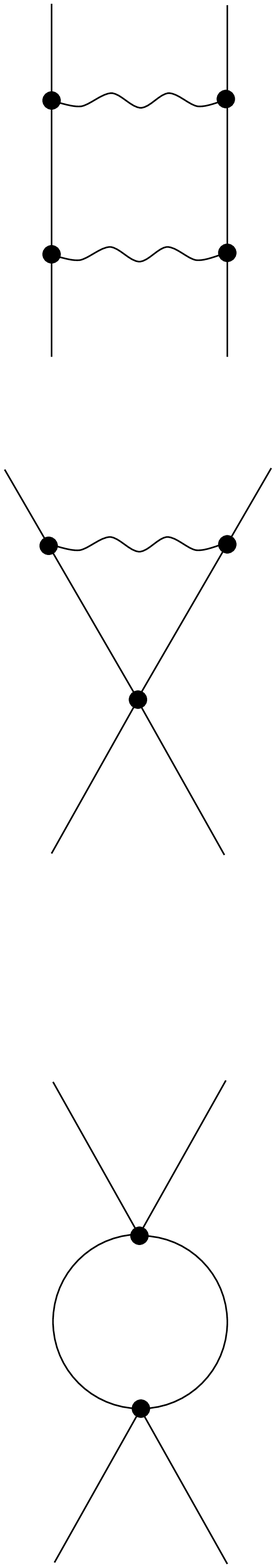}}
\vskip 0.5truecm
\caption{Feynman diagrams for the dynamic perturbation expansion of
$\Gamma_{04}$ to one--loop order.}
\label{gamma_04}
\end{figure}

\end{mcols} 


\begin{references}
\bibitem{dangerous_irrelevant_note} A dangerous irrelevant
perturbation is one which is irrelevant at the critical fixed point,
but relevant at the ordered fixed point, and often leads to violations
of hyperscaling.
\end{references}
\end{document}